\documentclass[%
reprint,
superscriptaddress,
amsmath,amssymb,
aps,
]{revtex4-2}

\usepackage{graphicx}
\usepackage{dcolumn}
\usepackage{bm}
\usepackage{hyperref}
\usepackage{float}
\usepackage{tabularx}
\usepackage{setspace}
\usepackage{outlines}
\usepackage{appendix}
\usepackage{color}
\usepackage[most]{tcolorbox}
\usepackage{varwidth}
\usepackage{capt-of}
\tcbuselibrary{listings}
\usepackage{bbm}

\tcbset{enhanced,colback=white!5!white,colframe=white!25!black,fonttitle=\bfseries}

\usepackage{xr}
\makeatletter
\newcommand*{\addFileDependency}[1]{
	\typeout{(#1)}
	\@addtofilelist{#1}
	\IfFileExists{#1}{}{\typeout{No file #1.}}
}
\makeatother
\newcommand*{\myexternaldocument}[1]{%
	\externaldocument{#1}%
	\addFileDependency{#1.tex}%
	\addFileDependency{#1.aux}%
}
\myexternaldocument{SI}

\newcommand{\bz}{\mathbf{z}}       
\newcommand{\bq}{\mathbf{q}}       
\newcommand{\bmu}{\boldsymbol{\mu}}       
\newcommand{\by}{\mathbf{y}}       

\usepackage{appendix}

\begin{document}
	
	\title{In-context learning emerges in chemical reaction networks without attention}
	
	\author{Carlos Floyd$^\S$}
	\email{csfloyd@uchicago.edu}
	\affiliation{Department of Chemistry, The University of Chicago, Chicago, Illinois 60637, USA}
	\affiliation{The James Franck Institute, The University of Chicago, Chicago, Illinois 60637, USA}
	\altaffiliation{These authors contributed equally to this work.}
	
	\author{Hector Manuel Lopez Rios$^\S$}
	\email{hectorlr@uchicago.edu}
	\affiliation{Department of Chemistry, The University of Chicago, Chicago, Illinois 60637, USA}
	\affiliation{The James Franck Institute, The University of Chicago, Chicago, Illinois 60637, USA}

	\author{Aaron R.\ Dinner}
	\affiliation{Department of Chemistry, The University of Chicago, Chicago, Illinois 60637, USA}
	\affiliation{The James Franck Institute, The University of Chicago, Chicago, Illinois 60637, USA}
	
	\author{Suriyanarayanan Vaikuntanathan}
	\email{svaikunt@uchicago.edu}
	\affiliation{Department of Chemistry, The University of Chicago, Chicago, Illinois 60637, USA}
	\affiliation{The James Franck Institute, The University of Chicago, Chicago, Illinois 60637, USA}

	\date{\today}
	\begin{abstract}
		We investigate whether chemical processes can perform in-context learning (ICL), a mode of computation typically associated with transformer architectures. ICL allows a system to infer task-specific rules from a sequence of examples without relying solely on fixed parameters. Traditional ICL relies on a pairwise attention mechanism which is not obviously implementable in chemical systems. However, we show theoretically and numerically that chemical processes can achieve ICL through a mechanism we call subspace projection, in which the entire input vector is mapped onto comparison subspaces, with the dominant projection determining the computational output. We illustrate this mechanism analytically in small chemical systems and show numerically that performance is robust to input encoding and dynamical choices, with the number of tunable degrees of freedom in the input encoding as a key limitation. Our results provide a blueprint for realizing ICL in chemical or other physical media and suggest new directions for designing adaptive synthetic chemical systems and understanding possible biological computation in cells.
		
	\end{abstract}
	
	\maketitle
	

	\section{Introduction}

\begin{figure*}[ht!]
		\begin{center}
			\includegraphics[width=\textwidth]{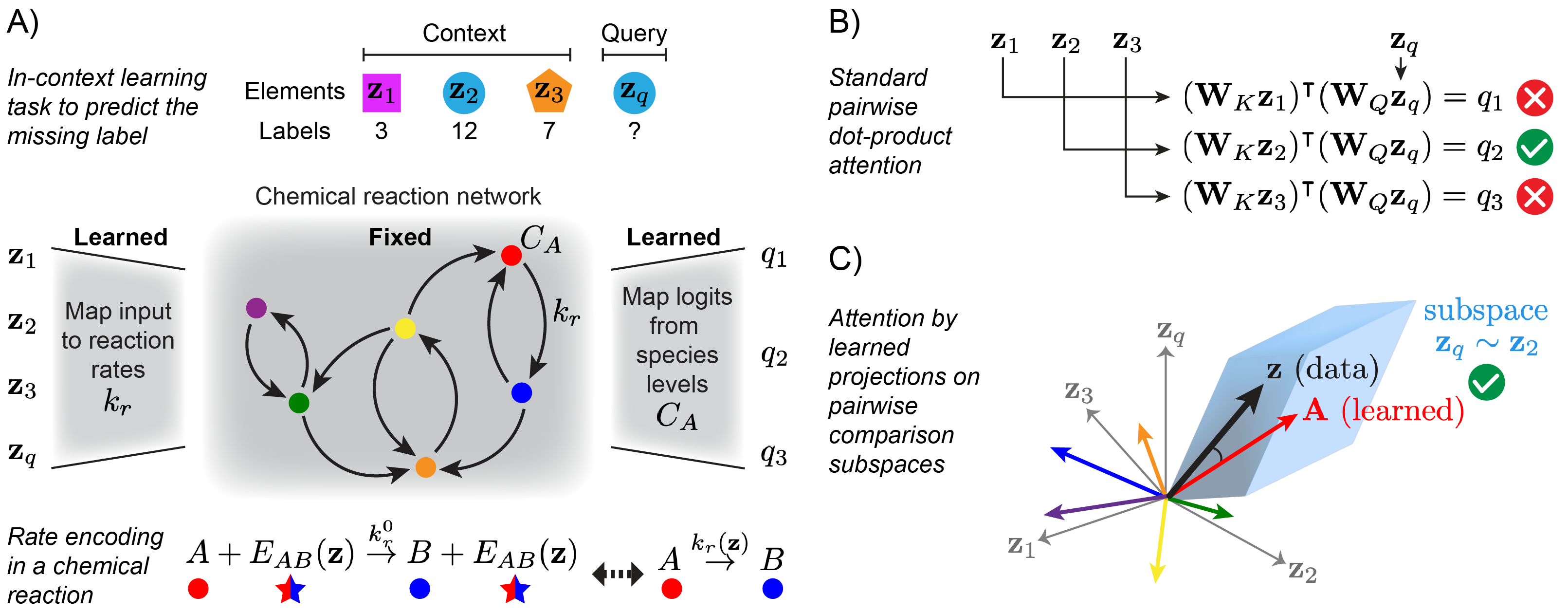}
			\caption{Schematic illustration of ICL by a chemical reaction network. 
				A)  Definition of the ICL task as matching a query element to an example within the context vector.  These context elements $\mathbf{z}_i$ are mapped into label predictions $q_i$ by the chemical reaction network.  
				B)  The standard mechanism for ICL uses a pairwise dot-product attention (involving learnable key $\mathbf{W}_K$ and query $\mathbf{W}_Q$ matrices) between the query element and the members of the context.  
				C)  Schematic illustration of the subspace projection mechanism, which involves projections of the context vector $\mathbf{z}$ onto different vectors which determine the steady-state output of the chemical reaction network (such as $\mathbf{A}$).  Training the network involves placing these vectors so that steady-state outputs can be decoded to solve the ICL task.  This is accomplished by placing these learnable vectors near certain subspaces which correspond to pairwise attention comparisons with the query component. }
			\label{ICLSchematic}
		\end{center}
	\end{figure*}

	Large language models (LLMs) represent a transformative class of machine learning systems with the striking ability to generalize beyond what is explicitly encoded in their parameters. Traditional models typically solve tasks through an ``in-weights learning'' (IWL) mechanism, in which the ability of the model to give accurate responses to queries depends on whether that type of query had been encountered during training and thus encoded in the model weights.  In contrast to IWL, ``in-context learning'' (ICL) refers to a mode of computation first characterized in transformer architectures (such as those used in LLMs), in which new task information can be supplied at query time through a sequence of input–output examples~\cite{brown2020language,dong2024survey,reddy2023mechanistic,nguyen2024differential,chan2022data}. The model then uses its internal activations to \emph{infer} a rule from these examples and apply that rule to new inputs, even when its weights were never explicitly trained on that type of input before (Figure~\ref{ICLSchematic}A). ICL therefore constitutes a form of on-the-fly or semantic learning, in which the correct response is extracted from the structure of the input context rather than solely from previously encountered examples.
	
	As part of a growing effort to implement computation in physical substrates beyond digital electronics~\cite{stern2023learning}, work during the past two decades has advanced synthetic chemical computing platforms including DNA circuits~\cite{vasle2024synthetic,seelig2006enzyme,qian2011neural,cherry2018scaling,cherry2025supervised,okumura2022nonlinear}, chemical reservoir computers~\cite{baltussen2024chemical,ghosh2025recursive}, and systems based on competitive association~\cite{evans2024pattern,chen2024synthetic} or programmable enzymatic networks~\cite{gao2018programmable,yang2025engineering}. These efforts demonstrate that chemical networks can support sophisticated nonlinear information processing by leveraging resource competition, interaction multiplicity, and nonequilibrium driving~\cite{hjelmfelt1991chemical,maass2000computational,genot2012computing,zhong2017associative,klumpe2023computational,parres2025contextual,floyd2025limits}, capabilities likely exploited in cellular decision-making and signaling~\cite{su2022ligand,lim2024cell}. However, such studies have so far focused only on IWL tasks of classification, pattern recognition, regression, and forecasting~\cite{floyd2025limits,evans2024pattern,kieffer2023molecular,dack2024recurrent,baltussen2024chemical,ghosh2025recursive}.
	
	Here we ask if it is possible for chemical reaction networks (CRNs) to exhibit ICL. An apparent challenge is that the core architectural motif underlying ICL in transformers is the pairwise dot-product attention mechanism \cite{bahdanau2014neural}, in which each element of the context interacts with the query through a learned similarity computation (Figure \ref{ICLSchematic}B).  Although synthetic chemical computers can be engineered through modular circuit design and, in principle, are capable of any computational task \cite{hjelmfelt1991chemical, qian2011neural}, we consider more general disordered chemical reaction networks without modular design or engineering.   Explicit dot-product attention is not obviously realizable in such disordered chemical systems, in which inputs (such as externally fixed chemical concentrations) are in general operated on by the whole chemical system rather than through pairwise comparisons of sub-networks. If disordered chemical systems, without any detailed engineering, are nevertheless capable of supporting ICL, this capability must arise from alternative biochemical or dynamical mechanisms that effectively reproduce the comparison and selection operations required for contextual inference.

	Here, we show theoretically and numerically that CRNs can perform ICL without relying on explicit pairwise attention. Instead, ICL emerges through a geometric mechanism which we call \emph{subspace projection} (Figure \ref{ICLSchematic}C). Unlike dot-product attention, this mechanism does not require an imposed procedure of decomposing the input into individual context elements; instead it naturally emerges during training.  We explicitly demonstrate how this mechanism operates in minimal chemical systems and show empirically that it is robust to different input encodings and choices of the chemical dynamics. Our results provide a framework for embedding ICL in chemical and other physical platforms, creating avenues for engineered adaptive chemistry and for elucidating possible computational principles used in biology.

	\section{Formulation of ICL for CRNs}
	
	To study ICL, we formulate the following classification task modeled along tasks in Ref.~\citenum{reddy2023mechanistic}. We consider a vector of elements $\bz \equiv \left(\bz_1,\, \bz_2,\, \ldots,\, \bz_{q}\right)$ of length $N_\text{c}+1$, where the first $N_\text{c}$ elements form the context and the last is the query.   Each $\bz_i\in\mathbb{R}^D$ is drawn from one of $L$ Gaussian mixture model (GMM) classes, with a within-class variation measure $\epsilon$.  Each class (or mode) of the GMM is equally probable, and the GMM data model is used to control the statistics of $\bz_i$ in this synthetic data; in practice $\bz_i$ typically correspond to latent vectors representing, for instance, word embeddings.  Each of the first $N_{\rm c}$ elements of this vector has a label $\ell_i$ assigned to the corresponding GMM class.  The last item at position $N_\text{c}+1$ is the so-called query item, whose label we want to predict. The vectors $\bz$ are engineered so that the query item is similar to one of the elements in the  conext. A system achieves successful ICL when it learns to predict the label corresponding to the context element with the highest overlap, even in cases where the context and labels are from GMM classes never seen or encountered in testing.

	
	We consider a chemical reaction network (CRN) with reactions rates $k_r$ (with $r$ labeling the reaction index).  We couple these rates to the context vector $\mathbf{z}$ via functions of the form 
	\begin{equation}
		k_r = \sigma\left(b_r + \mathbf{K}_r^\intercal\bz \right). \label{eqkr}
	\end{equation}
	Here $\sigma$ is an element-wise operation to ensure positivity of the reaction rate.  For example, one can imagine a reaction $A\rightarrow B$ catalyzed by an enzyme $E_{AB}$ whose reaction is externally fixed and coupled to $\mathbf{z}$ (Figure \ref{ICLSchematic}A).  The vector $\mathbf{K}_r \in \mathbb{R}^{(N_\text{c}+1)D}$ and the scalar $b_r$ are learned parameters for the map from the context vector $\mathbf{z}$ into the reaction rate $k_r$.  
	
	\begin{figure*}[ht!]
		\begin{center}
			\includegraphics[width=\textwidth]{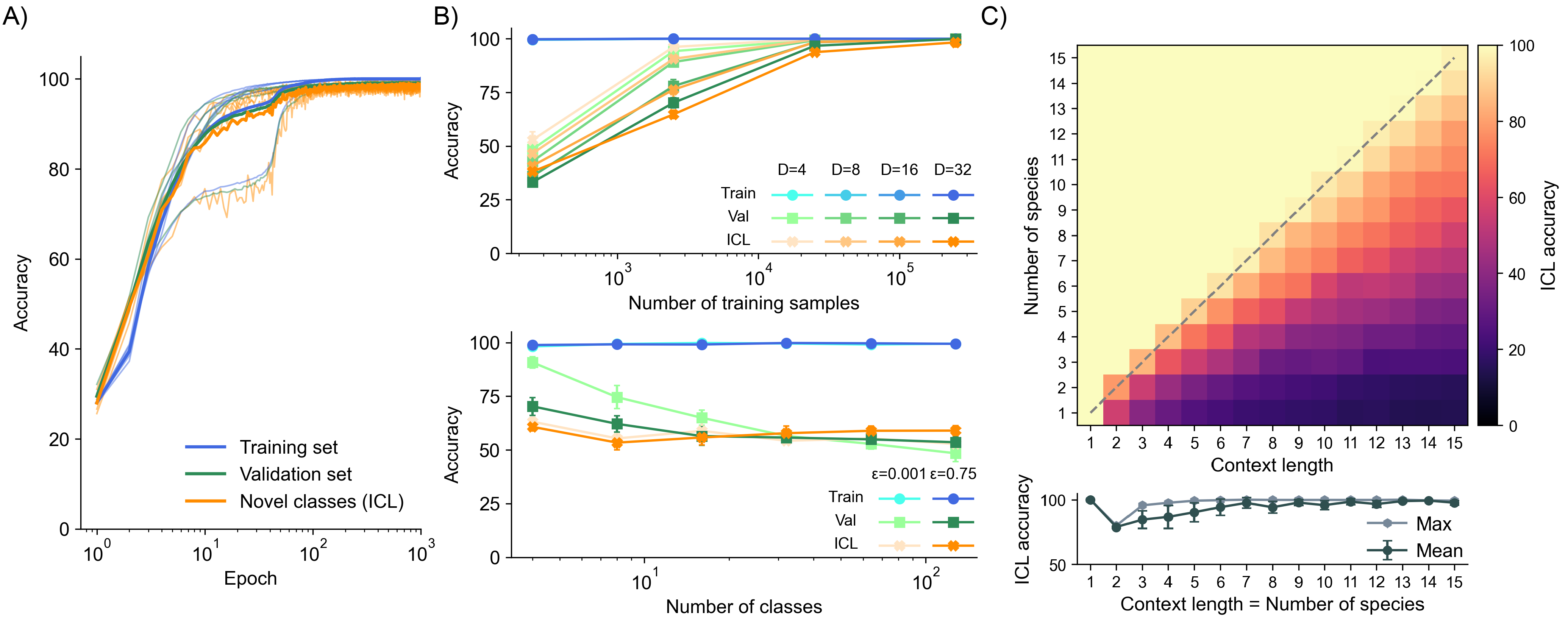}
			\caption{ICL performance across training conditions.  
				A) Plots of the training and validation set accuracy as well as accuracy on novel classes (termed ICL accuracy) as training progress.  Five runs and their average are shown for an example with context element dimension $D = 8$; the default parameters (see Supplementary Material Section \ref{SIsec:params}) are used throughout the paper unless otherwise specified.   
				B) Plots of the different accuracy measures after training networks across different conditions, with the mean and standard deviation over five runs shown for each condition.  In the top plot, the number of training samples $N_\text{samp}$ and $D$ are varied together, while in the bottom the number of GMM classes $N_\text{class}$ and within-class variation $\epsilon$ are varied.  In the bottom plot we set $N_\text{samp} = 250$ to show the behavior with low training data.  
				C)  Heat map of the average ICL accuracy over five runs as the context length $N_\text{c}$ and number of species $N_\text{n}$ are varied, using $N_\text{samp} = 2.5 \times 10^5$.  The dashed gray line is used to visualize the values $N_\text{n} = N_\text{c}$.  The bottom panel shows along this diagonal the mean and standard deviation as well as the maximum of the ICL accuracy over five runs.  Note that classification with $N_\text{c} =1$ is trivial.
			}
			\label{TrainingResults}
		\end{center}
	\end{figure*}

	\begin{figure*}[ht!]
		\begin{center}
			\includegraphics[width=\textwidth]{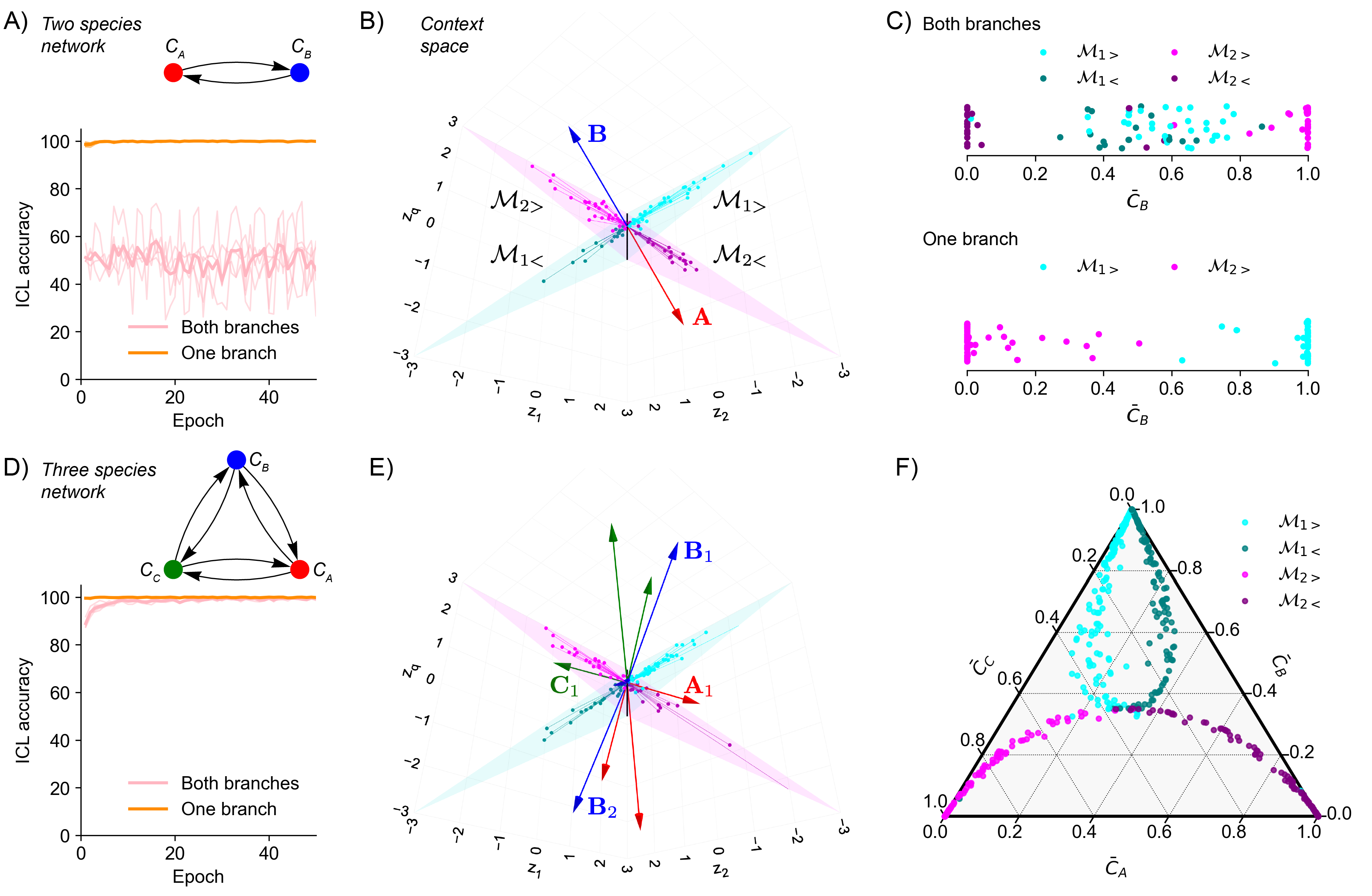}
			\caption{Demonstration of ICL mechanism in small networks.  
				A) For a two-species network, the ICL accuracy as training progresses ($N_{\mathrm{c}} = 2$, $D = 1$). Five trials are shown for each condition. In the ``one branch'' condition, networks are trained only on context vectors $\mathbf{z} = (z_1, z_2, z_q)$ satisfying $z_q = \max(z_1, z_2)$, corresponding to the subspaces $\mathcal{M}_{1>}$ and $\mathcal{M}_{2>}$. In the ``both branches'' condition, no restriction is applied and contexts are drawn from all subspaces.  
				B) Depiction of the context space of $\mathbf{z}$. Scatter points show context vectors sampled during training, colored according to the subspace branch they lie on. The linear subspaces $\mathcal{M}_{1}$ and $\mathcal{M}_{2}$ are shown as planes, with their intersection indicated by a solid black line. The learned vectors $\mathbf{A}$ and $\mathbf{B}$ that determine the steady state are shown as red and blue arrows (see main text for details); their lengths are not drawn to scale.
				C) Scatter plot of the steady-state concentration of species $B$ (here $C_{\text{tot}} = 1$) for context vectors sampled during training with one branch and with both branches. Small random vertical offsets are applied for visualization.  
				D) Same as panel A, but for a three-species network.  
				E) Same as panel B, but showing the additional learned vectors for the three-species network. See main text and Supplementary Material Section \ref{SIsec:threespecies} for details.  
				F) Ternary plot showing the steady-state distribution over $\bar{\mathbf{C}}$ with $C_{\text{tot}} = 1$ for context vectors sampled during training with both branches.  
			}
			\label{SmallNetworks}
		\end{center}
	\end{figure*}
	
	At steady state, the CRN has concentrations $\bar{\mathbf{C}} \in \mathbb{R}^{N_{\text{n}}}$ for its $N_{\text{n}}$ species, where the overbar denotes the steady-state value.  Classification is performed by first linearly mapping $\bar{\mathbf{C}}$ to vector of logits $\bq = \mathbf{B}\bar{\mathbf{C}}$ where $\mathbf{B} \in \mathbb{R}^{N_\text{c}\times N_\text{n}}$ is a learned matrix.  Class probabilities $\by \in \mathbb{R}^{N_\text{c}}$ are then determined through a softmax of $\bq$ with inverse temperature $1$:
	\begin{equation}
		y_i = \frac{
			\exp\left(q_{i}\right)}{
			\sum_{j=1}^{N_\text{c}} \exp\left(q_j\right). \label{eqsoftmax}
		}
	\end{equation} For a training example with target class $\ell^*$, the model output and loss are $\mathcal{L} = -\log y_{\ell^*}$. The label prediction corresponds to the index with the largest logit $(\mathbf{B}\bar{\mathbf{C}})_i$ while the softmax form enables standard negative–log-likelihood training (see Supplementary Material Section \ref{SIsec:Training} for details).
	
	We first consider chemical reactions described by first-order kinetics and a single conservation law, 
	Such first-order models are described by the dynamics
	\begin{equation}
		\dot{C}_n = \sum_{m\neq n}\left(k_{m\rightarrow n} C_m - k_{n\rightarrow m} C_n\right) \label{eqlindyn}
	\end{equation}
	together with the conservation law $\sum_n C_n = C_\text{tot}$.  If the reactions $m\rightarrow n$ form a connected graph then there is a unique steady state with an exact analytical expression given by the matrix-tree theorem \cite{schnakenberg1976network, owen2020universal}.   Although the dynamics in these systems are linear in the concentration variables, the steady-state concentrations are a non-linear functions of the rates and therefore of $\mathbf{z}$ \cite{floyd2025limits}.  
	
	In later sections we generalize our findings to chemical reaction networks to include non-linear kinetics and competitive winner-take-all models. In experiments, previous work consisting of classification~\cite{qian2011neural,cherry2025supervised} and winner-take-all computations~\cite{chen2024synthetic,cherry2018scaling} using chemical systems relied on the manipulation chemical species' concentrations. Usually, input and intermediary species concentrations, combined, determine downstream kinetics which lead to targeted outputs, thus enabling a form of computation, see Fig.~\ref{ICLSchematic}A) lower panel. In the specific case for winner-take-all reaction networks, Chen {\it et al.}~\cite{chen2024synthetic} experimentally constructed a $\mathbf{K}_r$-like matrix from partial concentrations of intermediary chemical species that bind with input chemical species concentrations. In Supplementary Material Section~\ref{SI:exp-wta} we propose how a winner-take-all chemical network under the experimental framework of \cite{chen2024synthetic} may be implemented that shows ICL.
	
	\section{ICL robustly emerges in CRNs across training conditions}
	We next empirically characterize ICL performance in first-order CRNs.  We distinguish three performance metrics: accuracy of class prediction evaluated on the training data, accuracy on a validation data set drawn from the same Gaussian mixture model (GMM) classes but withheld during training, and genuine ICL accuracy evaluated on data drawn from new GMM classes. See Supplementary Material Sections \ref{SIsec:GMM} -\ref{SIsec:params} for details and for the default parameters used throughout the paper. Figure \ref{TrainingResults}A shows typical trajectories of these accuracy measures, which saturate at close to 100\% as training progresses for several random seeds. This behavior confirms our proposal that linear chemical reaction network dynamics can produce steady states that support semantic classification over context vectors.
	
	We explore the performance of ICL as we vary several training and network hyperparameters (Figure \ref{TrainingResults}).  To understand the regimes in which ICL occurs, we first vary the context element dimension $D$ and the number of samples in the training data set used during supervised learning (Figure \ref{TrainingResults}B, top panel). We find that accuracy on the training set remains high for all conditions, but with insufficient training data the learned parameters overfit, leading to poor validation and ICL accuracy scores. This effect is exacerbated at higher $D$ because more parameters must be learned, requiring additional data to avoid overfitting. In the bottom panel of Figure \ref{TrainingResults}B, we vary the number of GMM classes in the training set together with the within-class variation parameter $\epsilon$. For few classes and low noise, both the training and validation accuracies remain relatively high while the ICL accuracy is low, indicating that the model has effectively memorized the GMM distribution on which it was trained and does not display true ICL, as it fails to make accurate predictions on data from unseen classes. Increasing the number of classes in this low-data regime does not improve ICL accuracy and instead degrades validation accuracy. Thus the appearance of overfitting can be masked by deceptively high validation accuracy when within-class variation is low and the number of classes is small because the data withheld during training looks sufficiently similar to the training data. To summarize, these results show that genuine ICL requires sufficient data complexity and diversity to prevent the model from memorizing class-specific features (an ``in-weights mechanism'') rather than the underlying generalizable rule (the ``semantic structure'').
	
	In Figure \ref{TrainingResults}C we vary the context length $N_\text{c}$ together with the number of species in the network $N_\text{n}$. For a linear readout with $\mathbf{B}$, we na\"{\i}vely expect that the context length cannot exceed the number of species; otherwise a single species would need to be decoded into more than one context-match prediction. We indeed find that the scaling $N_\text{c} \leq N_\text{n}$ approximately determines the condition for ICL feasibility when both $N_\text{n}$ and $N_\text{c}$ are large. We observe deviations from this prediction $N_\text{n} = N_\text{c} = 2$, however, indicating that there may also be a bottleneck on ICL performance arising from the ability to encode the context into the steady-state concentrations. We explore the origins of this bottleneck in the next section.  Finally, we verify that ICL is robust to the choice of rate-encoding function $\sigma$, obtaining similar results scaing results of ICL accuracy with $N_\text{n}$ and $N_\text{c}$ using alternative positive maps such as sigmoid and softplus in place of the exponential (see Supplementary Material Section \ref{SIsec:rateencode}).

	\section{Mechanism of ICL exploits geometry of the context space}\label{sec:iclmech}
	
	\begin{figure*}[ht!]
		\begin{center}
			\includegraphics[width=\textwidth]{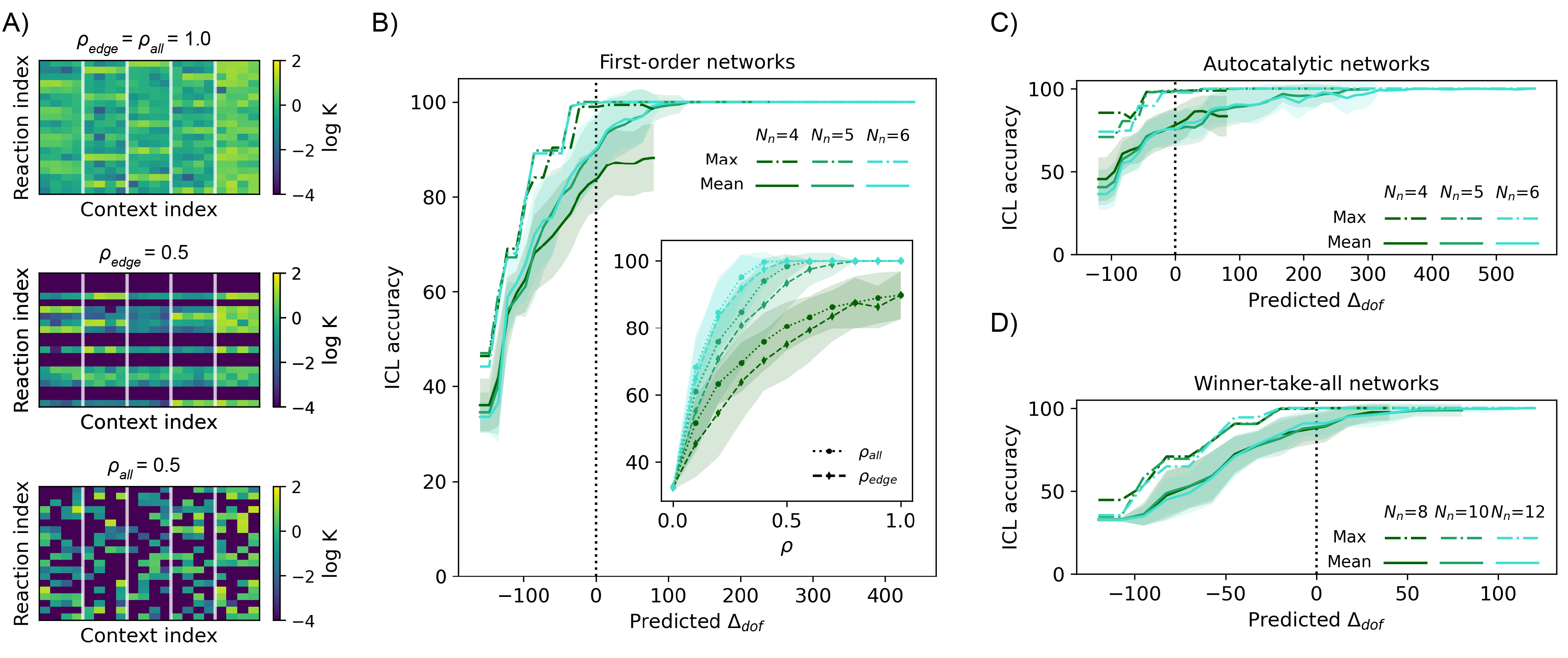}
			\caption{Sparsifying the input encoding across chemical models.    
				A)  The encoding vectors $\mathbf{K}_r$, shown horizontally for the 20 reactions in a trained fully connected linear reaction network with $N_\text{n} = 5$ species and context vectors with $N_\text{c} = 4$ and $D = 4$. White vertical lines separate the elements of the context vector. The corresponding encoding vectors trained after applying the sparsity masks $\rho_\text{edge}$ and $\rho_\text{all}$ are shown below.  
				B)  ICL accuracy for linear reaction networks of different sizes as $\rho_\text{all}$ is varied. For each value of $\rho_\text{all}$ and $N_\text{n}$, we run 30 trials with different random seeds controlling both sparsity-mask sampling and training stochasticity. We compute the number of degrees of freedom in the encoding vectors $\mathbf{K}_r$ and subtract $N_\text{req} = 2N_\text{c}(N_\text{c}+1)D$ to obtain the predicted difference $\Delta_\text{dof}$, which, when positive, indicates sufficient degrees of freedom to cover all ICL subspace branches. Running averages (solid lines), standard deviations (shaded regions), and maxima (dash–dot lines) are computed using windows of size 50. The inset shows data obtained using both $\rho_\text{all}$ and $\rho_\text{edge}$, plotting ICL accuracy directly against the sparsity parameter rather than against the sampled number of degrees of freedom.  Here the mean and standard deviation over the 30 runs for each parameter value are shown.
				C,D)  Same as the main plot of panel B, but for the autocatalytic and winner-take-all network models, respectively. Results using the $\rho_\text{edge}$ sparsity mask are shown in Supplementary Material Section \ref{SIsec:sparsity}.
			}
			\label{SparsityResults}
		\end{center}
	\end{figure*}
	To accomplish ICL, a CRN needs a modality to determine which of the input context elements ${\bf z}_i$ has high similarity with the query element, $\mathbf{z}_i \sim \mathbf{z}_{q}$ (with $\sim$ denoting high similarity). Because the CRN cannot form explicit pairwise dot products between $\mathbf{z}_{q}$ and the context elements $\mathbf{z}_i$, it must instead rely on a conceptually different mechanism. We propose and verify a geometric mechanism that enables ICL in CRNs. Importantly, this geometric approach shows how ICL can occur even in disordered systems without the need for explicit engineering. 
	
	We introduce the pairwise comparison subspaces
	\begin{equation}
		\mathcal{M}_i \equiv \left\{\mathbf{z} \,\big|\, \mathbf{z}_i \sim \mathbf{z}_{q}\right\} \label{eqMi}
	\end{equation}
	which comprises all inputs $\mathbf{z}$ in which the $i^\text{th}$ element has high similarity with the query element. If all inputs $\mathbf{z} \in \mathcal{M}_i$ can be ``classified'' by yielding a reliable output $\bar{\mathbf{C}}(\bz)$ through the learned projections, then this output can be decoded as the message ``$\mathbf{z}_i$ is similar to $\mathbf{z}_{q}$.''  Thus, similarity is determined by projections corresponding to subspaces that read out specific comparisons with context elements, rather than by directly evaluating the magnitude of each pairwise similarity.    
	
	We illustrate this concept of learned projection onto pairwise comparison subspaces in a small, analytically tractable example with $N_\text{n} = N_\text{c} = 2$, $D = 1$, and $b_r = 0$ for each reaction.  For concreteness, in the remainder of this paper we use a simple notion of similarity: the vectors $\mathbf{z}_i$ and $\mathbf{z}_{q}$ are considered similar when they are approximately equal
	\footnote{For moderately high-dimensional $\mathbf{z}_i$ that are randomly distributed in latent space with comparable norms, this notion of similarity in the standard attention model corresponds to choosing $\mathbf{W}_K^\intercal \mathbf{W}_Q = \mathbf{I}$, so that a standard dot product between $\mathbf{z}_{q}$ and $\mathbf{z}_i$ directly computes the similarity score $q_i$. The set-building condition $\mathbf{z}_i \sim \mathbf{z}_{q}$ in Equation~\ref{eqMi} might be replaced with conditions such as $\mathbf{W}_K \mathbf{z}_i \propto \mathbf{W}_Q \mathbf{z}_{q}$ to capture generalized notions of similarity, but we do not consider this further.}.  
	Using the matrix-tree theorem expression \cite{schnakenberg1976network} for the steady state of this reaction network with species $A$ and $B$ gives $\bar{C}_A = Z^{-1} \exp\left(\mathbf{A}^\intercal \mathbf{z}\right)$ and $\bar{C}_B = Z^{-1} \exp\left(\mathbf{B}^\intercal \mathbf{z}\right)$ with $Z = \exp\left(\mathbf{A}^\intercal \mathbf{z}\right) + \exp\left(\mathbf{B}^\intercal\mathbf{z}\right)$.  The two learned encoding vectors $\mathbf{A} = \mathbf{K}_1$ and $\mathbf{B} = \mathbf{K}_2 \in \mathbb{R}^{3}$ determine how the context vector is mapped into reaction space.
	
	For perfect ICL, $\mathbf{A}$ and $\mathbf{B}$ must separate all context vectors so that $\mathbf{z}\in\mathcal{M}_1$ maps to, for instance, dominant $\bar{C}_A$ and $\mathbf{z}\in\mathcal{M}_2$ to dominant $\bar{C}_B$. The intersection $\mathcal{M}_1 \cap \mathcal{M}_2$ splits each subspace into two branches $\mathcal{M}_1 = \mathcal{M}_1^> \cup \mathcal{M}_1^<$ and $\mathcal{M}_2 = \mathcal{M}_2^> \cup \mathcal{M}_2^<$, the branches being distinguished by whether $z_q = \max(z_1, z_2)$ or $z_q = \min(z_1, z_2)$ or more generally by the sign of the projection of a representative vector with $\mathbf{z}$.  With only two vectors it is impossible to cover all four branches, so full separation is impossible and accuracy plateaus at $\sim 50\%$ (Figure~\ref{SmallNetworks}A). The trained networks reveal that $\mathcal{M}_1$ and $\mathcal{M}_2$ contain branches that cannot both be distinguished with two degrees of freedom (Figure~\ref{SmallNetworks}B–C); we elaborate on this in Supplementary Material Section \ref{SIsec:twospecies}  However, when training is restricted to a single branch per subspace, the two-species network succeeds.
	
	By contrast, increasing the number of species to $N_\text{n} = 3$ yields a steady state involving $9$ encoding vectors $\{\mathbf{A}_i, \mathbf{B}_i, \mathbf{C}_i\}_{i=1}^3$, which are linear combinations of 6 free learned vectors $\{\mathbf{K}_r\}_{r=1}^6$ (see Supplementary Material Section \ref{SIsec:threespecies} for details).  This increase in learnable degrees of freedom allows the trained network to form distinct vectors with dominant overlaps for all $\mathbf{z} \in \mathcal{M}_1$ and $\mathbf{z} \in \mathcal{M}_2$ without restricting to a single branch per subspace (Figures~\ref{SmallNetworks}D,E).  The resulting steady-state response supports linear decoding of the predicted match with the context element (Figure~\ref{SmallNetworks}F).
	
	\section{Encoding sparsity predictably tunes ICL accuracy}\label{sec:sparsity}
	
	We see that the ICL task in CRNs requires positioning a set of encoding vectors within the context space such that they have dominant overlaps, and corresponding steady-state outcomes, with context vectors $\mathbf{z}$ belonging to the pairwise comparison subspaces $\mathcal{M}_i$. A fundamental limitation on a network’s ability to perform ICL is therefore the number of learnable degrees of freedom available to position these encoding vectors. An approximate estimate of the number of degrees of freedom required to uniquely cover both branches of each pairwise comparison subspace is $n_\text{req} = 2N_\text{c}(N_\text{c}+1)D$, since there are $2N_\text{c}$ such subspaces and each vector in the context space has dimension $(N_\text{c}+1)D$.
	
	In fully connected first-order reaction networks with $N_\text{n}$ species, there are $N_\text{n}(N_\text{n}-1)$ directed reaction edges, each involving a learned vector $\mathbf{K}_r \in \mathbb{R}^{(N_\text{c}+1)D}$ that is projected onto the context vector $\mathbf{z}$. Such networks therefore possess 
	$n_\text{g} = N_\text{n}(N_\text{n}-1)(N_\text{c}+1)D$ learnable degrees of freedom. Comparing $n_\text{g}$ to the required number of degrees of freedom $n_\text{req}$ for different values of $N_\text{n}=N_\text{c}$ reveals that only for sufficiently large networks, specifically $N_\text{n} \geq 3$, does $n_\text{g} > n_\text{req}$. For even larger $N_\text{c}$, the dominant bottleneck is therefore not the network’s ability to encode information in the reaction rates, but rather the ability of the decoder to assign $N_\text{c}$ prediction outcomes from $N_\text{n}$ steady-state species concentrations using a linear readout mechanism. This explains the observed accuracy trends in Figure~\ref{TrainingResults}C.
	
	We first extend this argument by examining how sparsifying the encoding operation affects the resulting accuracy. We implement sparsity in two ways: by applying a mask that decouples entire reaction edges from the input vector with probability $\rho_\text{edge}$, and by applying a second mask that decouples element-wise connections between individual reactions and components of the context vector with probability $\rho_\text{all}$ (see Figure~\ref{SparsityResults}A).  
	
	The results of varying $\rho_\text{edge}$ and $\rho_\text{all}$ for the first-order chemical model is shown in Figures~\ref{SparsityResults}B and in Supplementary Material Section \ref{SIsec:sparsity}. We plot the ICL accuracy after training against $\Delta_\text{dof}$, defined as the difference between the number of degrees of freedom counted in the set of encoding vectors $\mathbf{K}_r$ for a given sparsity realization and the required number $n_\text{req}$. When $\Delta_\text{dof} > 0$, our simple estimate that sufficient degrees of freedom are available to cover all branches of the pairwise comparison subspaces suggests that training to perfect accuracy should be possible. We find that this estimate works surprisingly well in describing the maximum accuracy achieved across different randomly initialized training runs for several system sizes. The mean accuracy across these samples is also approximately captured by the condition $\Delta_\text{dof} > 0$, with training becoming progressively easier, in the sense of higher mean accuracies, as $\Delta_\text{dof}$ increases further. We therefore conclude that $n_\text{req}$ provides a useful estimate of the number of degrees of freedom required to adequately cover the pairwise comparison subspaces for successful encoding.
	
	We also observe that higher typical accuracies are achieved when sparsifying the network using the $\rho_\text{all}$ mask than when using the $\rho_\text{edge}$ mask. This is likely because removing entire reaction couplings eliminates $\mathbf{K}_r$ vectors altogether, whereas under the $\rho_\text{all}$ sparsity pattern the number of encoding vectors typically remains unchanged, with sparsity instead restricting their ability to be positioned freely within the context space. Our previous work (Ref.~\citenum{floyd2025limits}) suggests that the number of reactions affected per input variable plays a central role in determining the computational expressivity of reaction networks, consistent with the observed numerical trends.
	
	\section{Higher order chemical reactions can achieve ICL more robustly and with smaller system sizes} \label{sec:nonlinmodels}
	To explore the generality of ICL beyond linear reaction networks, we consider two classes of non-linear chemical dynamics (model details are provided in Supplementary Material Section \ref{SIsec:models}). The first is an autocatalytic model incorporating second-order reactions $m + l \rightarrow n + l$, in which species number $l$ catalyzes the conversion of $m$ to $n$. We assume collective conservation of all species, $\sum_n C_n(t) = C_{\text{tot}}$. All possible autocatalytic reactions are allowed, with the corresponding second-order rates coupled to the context vector $\mathbf{z}$. The second model is the winner-take-all system described in Ref.~\citenum{genot2012computing}, in which species undergo opposite processes of degradation and production, with production involving competition for a shared resource. Concentrations are not conserved, and the species that captures the dominant share of the resource grows while others decay to extinction.   
	
	We study how the degree of freedom counting argument holds in these  chemical models with non-linear kinetics.  The results in Figures~\ref{SparsityResults}C,D show that the typical statistical behavior is well-described by the counting argument even in these more complex models.  Interestingly, however, we find that it is possible for non-linear networks to access more efficient encoding schemes using operations like thresholding. In Supplementary Material Section \ref{SIsec:wta} we identify a mechanism by which winner-take-all dynamics can circumvent the need to cover all pairwise comparison subspaces, instead assigning certain ICL outcomes by default when no context vectors have sufficiently large projections onto the encoding vectors $\mathbf{K}_r$. This highly nonlinear, competition-based mechanism allows winner-take-all networks to achieve accurate ICL with fewer degrees of freedom than predicted by $\Delta_\text{dof}$. A more thorough comparison of the computational expressivity tradeoffs between these different chemical dynamics is left for future work.

	\section{Conclusion}
	We have shown that in-context learning is a robust capability of chemical computation, arising naturally in simple reaction network architectures without network engineering. Across a range of models and training conditions, the primary bottleneck for successful ICL is not the specific form of the dynamics or encoding method, but rather the availability of sufficient degrees of freedom to cover the relevant pairwise comparison subspaces.
	In Section \ref{sec:iclmech} we analyzed the mechanism of ICL in the linear setting with exponential rate encoding. While the corresponding mechanism in non-linear systems is harder to interpret directly, the mechanism these networks may be using might be connected with findings such as those in Ref ~\cite{tong2024mlps} where MLP architectures were also shown to perform ICL like tasks. More generally they be viewed as reservoir-like dynamical processors~\cite{nakajima2021reservoir}, where high-dimensional and somewhat arbitrary chemical dynamics are likewise exploited for computation provided encoding and decoding capabilities. Recent implementations of chemical reservoir computers show that, given suitable encoding and decoding schemes, highly complex internal dynamics can support reliable IWL computation~\cite{baltussen2024chemical,ghosh2025recursive}. Our results expand the class of tasks that chemical reservoir computers can perform to include ICL.  
	
	The effective mechanism of attention that emerges in these systems differs qualitatively from the explicit pairwise dot-product attention used in transformer architectures. Instead, comparison and selection are implemented through projections of the entire context vector onto learned vectors that determine steady-state outcomes, with ICL accuracy governed by the ability of these vectors to span different pairwise comparison subspaces. While some treatments of linear attention architectures can be formulated as operations on the entire context vector, they can typically still be decomposed into effective pairwise operations \cite{lu2025asymptotic}. Our results therefore highlight a distinct, intrinsically global form of attention based on subspace coverage rather than explicit similarity computation.
	
	Finally, our framework raises broader questions about how contextual information might be encoded in real chemical or biological systems, and more generally in physical dynamical systems. While we focused on chemical reaction networks with input-dependent rates, many of our arguments rely only on the structure of the dynamics and the dimensionality of their controllable parameters. This suggests that a wide class of dynamical systems, not limited to CRNs, may be capable of supporting ICL-like behavior when appropriately parameterized. For example, spring or resistor networks trained for in-weight classification tasks may also be capable of in-context classification \cite{stern2021supervised, stern2023learning}. Exploring how such mechanisms could be realized in biological regulation, synthetic chemical circuits, or other physical substrates remains a direction for future work.

	\section*{Acknowledgments}
	This work was supported by DOE BES Grant DE-SC0019765 to SV. We gratefully acknowledge discussions with Gautam Reddy and Akshit Goyal. 
	We acknowledge support from the National Science Foundation through the Physics Frontier Center for Living Systems (PHY-2317138)
	CF acknowledges support from the University of Chicago Data Science Institute AI + Science Research Initiative.  The authors acknowledge the University of Chicago’s Research Computing Center for computing resources.

	\bibliographystyle{unsrt}
	\bibliography{CLCRN}

@article{tong2024mlps,
  title={MLPs learn in-context on regression and classification tasks},
  author={Tong, William L and Pehlevan, Cengiz},
  journal={arXiv preprint arXiv:2405.15618},
  year={2024}
}

@article{floyd2025limits,
  title={Limits on the computational expressivity of non-equilibrium biophysical processes},
  author={Floyd, Carlos and Dinner, Aaron R and Murugan, Arvind and Vaikuntanathan, Suriyanarayanan},
  journal={Nature Communications},
  volume={16},
  number={1},
  pages={7184},
  year={2025},
  publisher={Nature Publishing Group UK London}
}

@article{seelig2006enzyme,
  title={Enzyme-free nucleic acid logic circuits},
  author={Seelig, Georg and Soloveichik, David and Zhang, David Yu and Winfree, Erik},
  journal={science},
  volume={314},
  number={5805},
  pages={1585--1588},
  year={2006},
  publisher={American Association for the Advancement of Science}
}

@article{dack2024recurrent,
  title={Recurrent neural chemical reaction networks that approximate arbitrary dynamics},
  author={Dack, Alexander and Qureshi, Benjamin and Ouldridge, Thomas E and Plesa, Tomislav},
  journal={arXiv preprint arXiv:2406.03456},
  year={2024}
}

@article{kieffer2023molecular,
  title={Molecular computation for molecular classification},
  author={Kieffer, Coline and Genot, Anthony J and Rondelez, Yannick and Gines, Guillaume},
  journal={Advanced Biology},
  volume={7},
  number={3},
  pages={2200203},
  year={2023},
  publisher={Wiley Online Library}
}

@article{evans2024pattern,
  title={Pattern recognition in the nucleation kinetics of non-equilibrium self-assembly},
  author={Evans, Constantine Glen and O’Brien, Jackson and Winfree, Erik and Murugan, Arvind},
  journal={Nature},
  volume={625},
  number={7995},
  pages={500--507},
  year={2024},
  publisher={Nature Publishing Group UK London}
}

@article{vasle2024synthetic,
  title={Synthetic biological neural networks: From current implementations to future perspectives},
  author={Vasle, Ana Halu{\v{z}}an and Mo{\v{s}}kon, Miha},
  journal={BioSystems},
  volume={237},
  pages={105164},
  year={2024},
  publisher={Elsevier}
}

@article{ghosh2025recursive,
  title={A recursive enzymatic competition network capable of multitask molecular information processing},
  author={Ghosh, Souvik and Baltussen, Mathieu G and Knox, Anna C and Haije, Rianne and Duez, Quentin and Tsitsimeli, Anastasia T and Chak, Man Him and Beves, Jonathon E and Huck, Wilhelm TS},
  journal={Nature Chemistry},
  pages={1--7},
  year={2025},
  publisher={Nature Publishing Group UK London}
}

@article{baltussen2024chemical,
  title={Chemical reservoir computation in a self-organizing reaction network},
  author={Baltussen, Mathieu G and de Jong, Thijs J and Duez, Quentin and Robinson, William E and Huck, Wilhelm TS},
  journal={Nature},
  volume={631},
  number={8021},
  pages={549--555},
  year={2024},
  publisher={Nature Publishing Group UK London}
}

@article{qian2011neural,
  title={Neural network computation with DNA strand displacement cascades},
  author={Qian, Lulu and Winfree, Erik and Bruck, Jehoshua},
  journal={nature},
  volume={475},
  number={7356},
  pages={368--372},
  year={2011},
  publisher={Nature Publishing Group UK London}
}

@article{cherry2025supervised,
  title={Supervised learning in DNA neural networks},
  author={Cherry, Kevin M and Qian, Lulu},
  journal={Nature},
  pages={1--9},
  year={2025},
  publisher={Nature Publishing Group UK London}
}

@article{yang2025engineering,
  title={Engineering synthetic phosphorylation signaling networks in human cells},
  author={Yang, Xiaoyu and Rocks, Jason W and Jiang, Kaiyi and Walters, Andrew J and Rai, Kshitij and Liu, Jing and Nguyen, Jason and Olson, Scott D and Mehta, Pankaj and Collins, James J and others},
  journal={Science},
  volume={387},
  number={6729},
  pages={74--81},
  year={2025},
  publisher={American Association for the Advancement of Science}
}

@article{chen2024synthetic,
  title={A synthetic protein-level neural network in mammalian cells},
  author={Chen, Zibo and Linton, James M and Xia, Shiyu and Fan, Xinwen and Yu, Dingchen and Wang, Jinglin and Zhu, Ronghui and Elowitz, Michael B},
  journal={Science},
  volume={386},
  number={6727},
  pages={1243--1250},
  year={2024},
  publisher={American Association for the Advancement of Science}
}

@article{okumura2022nonlinear,
  title={Nonlinear decision-making with enzymatic neural networks},
  author={Okumura, Shu and Gines, Guillaume and Lobato-Dauzier, Nicolas and Baccouche, Alexandre and Deteix, Robin and Fujii, Teruo and Rondelez, Yannick and Genot, Anthony J},
  journal={Nature},
  volume={610},
  number={7932},
  pages={496--501},
  year={2022},
  publisher={Nature Publishing Group UK London}
}

@article{cherry2018scaling,
  title={Scaling up molecular pattern recognition with DNA-based winner-take-all neural networks},
  author={Cherry, Kevin M and Qian, Lulu},
  journal={Nature},
  volume={559},
  number={7714},
  pages={370--376},
  year={2018},
  publisher={Nature Publishing Group UK London}
}

@article{parres2025contextual,
  title={Contextual computation by competitive protein dimerization networks},
  author={Parres-Gold, Jacob and Levine, Matthew and Emert, Benjamin and Stuart, Andrew and Elowitz, Michael B},
  journal={Cell},
  volume={188},
  number={7},
  pages={1984--2002},
  year={2025},
  publisher={Elsevier}
}

@article{hjelmfelt1991chemical,
  title={Chemical implementation of neural networks and Turing machines.},
  author={Hjelmfelt, Allen and Weinberger, Edward D and Ross, John},
  journal={Proceedings of the National Academy of Sciences},
  volume={88},
  number={24},
  pages={10983--10987},
  year={1991}
}

@article{su2022ligand,
  title={Ligand-receptor promiscuity enables cellular addressing},
  author={Su, Christina J and Murugan, Arvind and Linton, James M and Yeluri, Akshay and Bois, Justin and Klumpe, Heidi and Langley, Matthew A and Antebi, Yaron E and Elowitz, Michael B},
  journal={Cell systems},
  volume={13},
  number={5},
  pages={408--425},
  year={2022},
  publisher={Elsevier}
}

@article{klumpe2023computational,
  title={The computational capabilities of many-to-many protein interaction networks},
  author={Klumpe, Heidi E and Garcia-Ojalvo, Jordi and Elowitz, Michael B and Antebi, Yaron E},
  journal={Cell systems},
  volume={14},
  number={6},
  pages={430--446},
  year={2023},
  publisher={Elsevier}
}

@article{gao2018programmable,
  title={Programmable protein circuits in living cells},
  author={Gao, Xiaojing J and Chong, Lucy S and Kim, Matthew S and Elowitz, Michael B},
  journal={Science},
  volume={361},
  number={6408},
  pages={1252--1258},
  year={2018},
  publisher={American Association for the Advancement of Science}
}

@book{lim2024cell,
  title={Cell signaling: principles and mechanisms},
  author={Lim, Wendell A and Mayer, Bruce J},
  year={2024},
  publisher={CRC Press}
}

@article{zhong2017associative,
  title={Associative pattern recognition through macro-molecular self-assembly},
  author={Zhong, Weishun and Schwab, David J and Murugan, Arvind},
  journal={Journal of Statistical Physics},
  volume={167},
  number={3},
  pages={806--826},
  year={2017},
  publisher={Springer}
}

@article{maass2000computational,
  title={On the computational power of winner-take-all},
  author={Maass, Wolfgang},
  journal={Neural computation},
  volume={12},
  number={11},
  pages={2519--2535},
  year={2000},
  publisher={MIT Press}
}

@article{genot2012computing,
  title={Computing with competition in biochemical networks},
  author={Genot, Anthony J and Fujii, Teruo and Rondelez, Yannick},
  journal={Physical review letters},
  volume={109},
  number={20},
  pages={208102},
  year={2012},
  publisher={APS}
}

@article{brown2020language,
  title={Language models are few-shot learners},
  author={Brown, Tom and Mann, Benjamin and Ryder, Nick and Subbiah, Melanie and Kaplan, Jared D and Dhariwal, Prafulla and Neelakantan, Arvind and Shyam, Pranav and Sastry, Girish and Askell, Amanda and others},
  journal={Advances in neural information processing systems},
  volume={33},
  pages={1877--1901},
  year={2020}
}

@article{dong2024survey,
  title={A survey on in-context learning},
  author={Dong, Qingxiu and Li, Lei and Dai, Damai and Zheng, Ce and Ma, Jingyuan and Li, Rui and Xia, Heming and Xu, Jingjing and Wu, Zhiyong and Chang, Baobao and others},
  journal={Proceedings of the 2024 conference on empirical methods in natural language processing},
  pages={1107--1128},
  year={2024}
}

@article{reddy2023mechanistic,
  title={The mechanistic basis of data dependence and abrupt learning in an in-context classification task},
  author={Reddy, Gautam},
  journal={arXiv preprint arXiv:2312.03002},
  year={2023}
}

@article{bahdanau2014neural,
  title={Neural machine translation by jointly learning to align and translate},
  author={Bahdanau, Dzmitry and Cho, Kyunghyun and Bengio, Yoshua},
  journal={arXiv preprint arXiv:1409.0473},
  year={2014}
}

@article{nguyen2024differential,
  title={Differential learning kinetics govern the transition from memorization to generalization during in-context learning},
  author={Nguyen, Alex and Reddy, Gautam},
  journal={ArXiv},
  pages={arXiv--2412},
  year={2024}
}

@article{chan2022data,
  title={Data distributional properties drive emergent in-context learning in transformers},
  author={Chan, Stephanie and Santoro, Adam and Lampinen, Andrew and Wang, Jane and Singh, Aaditya and Richemond, Pierre and McClelland, James and Hill, Felix},
  journal={Advances in neural information processing systems},
  volume={35},
  pages={18878--18891},
  year={2022}
}

@article{schnakenberg1976network,
  title={Network theory of microscopic and macroscopic behavior of master equation systems},
  author={Schnakenberg, J{\"u}rgen},
  journal={Reviews of Modern physics},
  volume={48},
  number={4},
  pages={571},
  year={1976},
  publisher={APS}
}

@article{owen2020universal,
  title={Universal thermodynamic bounds on nonequilibrium response with biochemical applications},
  author={Owen, Jeremy A and Gingrich, Todd R and Horowitz, Jordan M},
  journal={Physical Review X},
  volume={10},
  number={1},
  pages={011066},
  year={2020},
  publisher={APS}
}

@article{lu2025asymptotic,
  title={Asymptotic theory of in-context learning by linear attention},
  author={Lu, Yue M and Letey, Mary and Zavatone-Veth, Jacob A and Maiti, Anindita and Pehlevan, Cengiz},
  journal={Proceedings of the National Academy of Sciences},
  volume={122},
  number={28},
  pages={e2502599122},
  year={2025},
  publisher={National Academy of Sciences}
}

@article{stern2021supervised,
  title={Supervised learning in physical networks: From machine learning to learning machines},
  author={Stern, Menachem and Hexner, Daniel and Rocks, Jason W and Liu, Andrea J},
  journal={Physical Review X},
  volume={11},
  number={2},
  pages={021045},
  year={2021},
  publisher={APS}
}

@article{stern2023learning,
  title={Learning without neurons in physical systems},
  author={Stern, Menachem and Murugan, Arvind},
  journal={Annual Review of Condensed Matter Physics},
  volume={14},
  number={1},
  pages={417--441},
  year={2023},
  publisher={Annual Reviews}
}

@book{nakajima2021reservoir,
  title={Reservoir computing},
  author={Nakajima, Kohei and Fischer, Ingo},
  year={2021},
  publisher={Springer}
}
	
	\clearpage
	
	\appendix

	\begin{onecolumngrid}

		\section{Supplementary methods}
		\subsection{Data generation via Gaussian Mixture Models}\label{SIsec:GMM}
		
		\subsubsection{Gaussian Mixture Model structure}
		Training and evaluation data are generated from a Gaussian Mixture Model (GMM) with $K$ classes in a $D$-dimensional feature space. The model associates these $K$ classes with $L$ discrete labels, where $L \leq K$ in general, allowing multiple classes to share the same label.  We set $L=K$ throughout for simplicity.  For each class $k \in \{1, \ldots, K\}$, we sample a class mean
		\begin{equation}
			\bmu_k \sim \mathcal{N}\left(\mathbf{0}, \frac{1}{D}\mathbf{I}\right),
		\end{equation}
		where the $1/\sqrt{D}$ scaling ensures that inter-class distances remain roughly constant as dimensionality increases. Each class is then randomly assigned a discrete label $\ell_k \in \{1, 2, \ldots, L\}$. Individual feature vectors are sampled from a class by adding within-class Gaussian noise:
		\begin{equation}
			\bz \sim \mathcal{N}\left(\bmu_k, \frac{\epsilon^2}{D}\mathbf{I}\right),
		\end{equation}
		where $\epsilon$ controls the amount of within-class variation.
		
		\subsubsection{In-context learning sequence construction}
		Each training example consists of a sequence of $N_\text{c}$ context items followed by a single query item. To construct a sequence, we randomly select $N_\text{c}$ classes from the GMM (with replacement), so that each class appears at most once in the context. For each selected class, we sample a feature vector with independent noise:
		
		\begin{align}
			\text{Context items: } &\quad \bz_i \sim \mathcal{N}\left(\bmu_{k_i}, \frac{\epsilon^2}{D}\mathbf{I}\right), \quad i = 1, \ldots, N_\text{c} \\
			\text{Context labels: } &\quad \ell_i = \ell_{k_i}
		\end{align}
		where $k_i \in \{1, \ldots, K\}$ is the class index for the $i$-th context item. The query item is then drawn from one of the classes already present in the context:
		\begin{equation}
			\bz_{N_\text{c}+1} = \bz_{j^*} \quad j^* \in \{k_1, \ldots, k_{N_\text{c}}\}
		\end{equation}
		where $j^*$ is chosen uniformly at random from the elements the context. The target label for the sequence is thus $\ell^* = \ell_{j^*}$. This construction ensures that the model has access to at least one labeled example of the query's class within the context sequence, enabling in-context learning.  A variation is possible in which $\bz_{N_\text{c}+1}$ is sampled from the same Gaussian class as that of the $j^*$ context element to account for within-class variation of the target, but for simplicity we instead directly copy the element itself as the query element.
		
		\subsubsection{Training, validation, and novel class testing}
		
		A critical distinction exists between training/validation data and the novel class testing regime:
		\begin{itemize}
			\item Training and validation sets: Sequences are constructed using the $K$ class means $\{\bmu_k\}$ sampled during GMM initialization. The model observes these class means repeatedly across many training examples, allowing it to potentially memorize the association between specific feature patterns and labels in its weights.
			
			\item Novel class test of true ICL: To evaluate genuine in-context learning, we generate test sequences using \emph{completely novel class means} $\{\tilde{\bmu}_k\}$ sampled fresh from the same Gaussian distribution. These means have never been seen during training. Crucially, the label space remains the same ($\ell \in \{1, \ldots, L\}$), and each novel class appears once in the context with its assigned label. The model must therefore learn the class-label mappings purely from the context sequence, without relying on memorized weight patterns.
		\end{itemize}
		This testing paradigm directly probes whether the model has developed true in-context learning capabilities — the ability to learn new input-output mappings from examples provided at inference time — rather than merely retrieving memorized associations from training.
		
		\subsection{Training algorithm}\label{SIsec:Training}
		
		The model parameters, the rate coupling vectors $\{\mathbf{K}_r\}$ and biases $\{b_r\}$, along with the output mapping matrix $\mathbf{B}$, are learned via stochastic gradient descent on labeled training sequences.  For simplicity we set $b_r = 0$ throughout.  Each training example consists of a context sequence of $N_\text{c}$ items with known labels, followed by a query item whose label serves as the supervision signal.
		
		We optimize using the Adam optimizer with a learning rate of $10^{-3}$ over $200$ epochs. Training proceeds in mini-batches provided by the data loader, though batch size is configurable depending on memory constraints and dataset size. To prevent gradient explosion during training, we apply gradient clipping with a maximum norm of $1.0$ before each parameter update.

		\subsection{Parameters}\label{SIsec:params}
		We use the default parameters listed in Table \ref{tab:parameters} unless otherwise specified.  We note that the attention softmax temperature (used in Equation \ref{eqsoftmax}) is only relevant during training, as when we evaluate accuracies we use an effective temperature of 0.    
		
		\begin{table}[h]
			\centering
			\caption{Model parameters and default values}
			\begin{tabular}{|c|l|c|}
				\hline
				\textbf{Symbol} & \textbf{Definition} & \textbf{Value} \\
				\hline
				
				\multicolumn{3}{|c|}{\textit{Network and context parameters}} \\
				\hline
				$N_\text{n}$ & Number of chemical species & 5 \\
				$N_\text{c}$ & Length of context vector & 4 \\
				$D$ & Dimension of context element $\mathbf{z}$ & 4 \\
				\hline
				
				\multicolumn{3}{|c|}{\textit{Winner-take-all model parameters}} \\
				\hline
				$R^0$ & Amount of shared resource & 5 \\
				$\tau_\text{sp}$ & Softplus temperature & 0.1 \\
				$\tau_\text{sm}$ & Softmin temperature & 0.01 \\
				\hline
				
				\multicolumn{3}{|c|}{\textit{GMM data parameters}} \\
				\hline
				$L$ & Number of GMM classes & 128 \\
				$\epsilon$ & Within-class variation of GMMs & $10^{-3}$ \\
				\hline
				
				\multicolumn{3}{|c|}{\textit{Training parameters}} \\
				\hline
				--- & Batch size during training & 50 \\
				--- & Number of training samples & 25{,}000 \\
				--- & Number of validation samples & 5{,}000 \\
				--- & ADAM learning rate & $2.5 \times 10^{-3}$ \\
				--- & Number of epochs & 1{,}000 \\
				--- & Softmax temperature & $1$ (linear, autocatalytic), $0.1$ (WTA) \\
				\hline
				
			\end{tabular}
			\label{tab:parameters}
		\end{table}

		\subsection{Chemical models}\label{SIsec:models}
		The ``chemistry'' of the system dictates the mapping from the set of reaction rates $\{k_r\}_{r = 1}^{N_\text{r}}$ to the steady-state concentration vector $\bar{\mathbf{C}}$.  We consider three classes of chemical dynamics: a linear reaction network, an autocatalytic reaction network, and a winner-take-all network.  
		
		\subsubsection{Linear reaction networks}\label{FirstOrderModel}
		We primarily study ICL in linear reaction networks described by the chemical dynamics of Equation~\ref{eqlindyn} in the main text.  These dynamics obey the conservation law $\sum_n C_n(t) = C_\text{tot}$.  The context vector is mapped into the reaction rates of the network via Equation~\ref{eqkr} of the main text.  To compute the steady-state concentration profiles $\bar{\mathbf{C}}$ we use a differentiable iterative solution of the linear system $\mathbf{k}(\mathbf{z})^\intercal\bar{\mathbf{C}} = \mathbf{0}$ constrained to $\sum_n\bar{C}_n = 1$ (we set $C_\text{tot} = 1$ throughout) with an implementation provided by Pytorch.  We optimize the parameters $\mathbf{K}_r$, $\mathbf{B}$, and optionally $b_r$ during training.  
		
		\subsubsection{Autocatalytic reaction networks}\label{AutocatalyticModel}
		In the autocatalytic network model we add additional second-order reactions onto the linear reaction networks in a way that still conserves the total concentration.  Specifically we consider catalytic reactions of the form $m + l \rightarrow n + l$, in which species $l$ facilitates conversion of $m$ to $n$ (such as through a catalyzed phosphorylation reaction).  The dynamics are 
		\begin{equation}
			\dot{C}_n
			=
			\sum_{m\neq n}\left(
			k_{m\rightarrow n} C_m
			-
			k_{n\rightarrow m} C_n
			\right)
			+
			\sum_{l,m\neq n}
			\left(
			k^\text{a}_{ml\rightarrow nl} C_m C_l
			-
			k^\text{a}_{nl\rightarrow ml} C_n C_l
			\right).
			\label{eqautodyn}
		\end{equation}
		In the second sum $l$ is arbitrary but $m\neq n $ prevents self-loops.  The context vector is encoded into the autocatalytic reaction rates $k^\text{a}_r$ as in the linear model, 
		\begin{equation}
			k^\text{a}_r = \sigma(b_r +\mathbf{K}_r^\intercal\mathbf{z}).
		\end{equation}
		The first order rates $k_r$ are not coupled to $\mathbf{z}$ but can have non-zero values.  In training this model we freeze the $b_r$ parameters during training (for reactions who have survived sparsification) and learn $k_r$ (first-order rates) the $\mathbf{K}_r$ vectors, and $\mathbf{B}$.  To compute the steady-state differentiably, we minimize the corresponding quadratic function of $\bar{\mathbf{C}}$ constrained to $\sum_n \bar{C}_n = 1$ using an implementation of Newton optimization provided by PyTorch.

		\subsubsection{Winner-take-all networks}\label{SIsec:WTAModel}
		In WTA networks, a reduced amount of chemical species are designed to subsist in the steady state. The prevailing number of chemical species depends on the number of resources in the chemical reaction network. Different sets of prevailing chemical species are favored through inputs which modulate the network reaction rates. Here, we show ICL in WTA networks is possible using inputs composed of context and query lengths.  
		
		We calculate the quasi-rates $f_j$ from the context vector $\mathbf{z}$
		\begin{equation}
			k_r = \frac{1}{K^\text{eq}_r}\text{Softplus}\left (  \mathbf{K}_r^\intercal\mathbf{z} \right) \label{wtaratekr}
		\end{equation}
		where $K^\text{eq}_r$ (to be distinguished from the encoding vectors $\mathbf{K}_r$) is an equilibrium constant.  Softplus is defined as 
		\begin{equation}
			\text{Softplus}(x) = \tau_\text{sp}^{-1}\log\left(1 + \exp\left(\tau_{\text{sp}} x\right) \right) \label{wtasoftplus}
		\end{equation}
		with $\tau_\text{sp}$ controlling sharpness near $x = 0$.
		Given these rates, and assuming linear degradation $\beta_j$ and catalytic reaction dynamics with a resource $R^0$, we compute the steady-state concentrations as
		\begin{equation}
			\bar{C}_r = K^\text{eq}_r \, \text{Softmin}\left( \frac{\beta_r}{k_r} \right ) \ \text{Softplus} \left ( \frac{R^0 k_r}{\beta_r} - 1 \right ).  \label{wtaCr}
		\end{equation}
		Here $\text{Softmin}(\{x_r\}) = \text{Softmax}(\{-x_r\})$ involves a temperature $\tau_\text{sm}$.  In this model we train for $\mathbf{K}_r$, $K^\text{eq}_r$, $\beta_r$, and $\mathbf{B}$, while $R^0$ is held constant.  The steady-state is computed numerically using PyTorch implementations of the Softmin and Softplus functions.  
		
		\subsection{Possible experimental ICL implementation using winner-take-all networks}\label{SI:exp-wta}
		
		In Ref.~\citenum{chen2024synthetic}, Chen  {\it et. al} construct a WTA chemical reaction network using de-novo heterodimers and biochemical components. Unlike previous theoretical work on WTA networks, the role of their inputs do not necessarily affect reaction rates, but rather enable the formation of complexes that, through a series of self-activating and inhibitory reactions,  enable stable steady-state concentrations of a single chemical species. 
		
		Specifically, inputs for the experimental WTA network consist of concentrations of different heterodimers $\mathbf{X} = (X_1,X_2,\dots,X_{N_\text{m}})$ where $X_i$ is the concentration of the $i$-th heterodimer. A heterodimer binds to an intermediary component forming a complex. Each of the $N_\text{n}$ intermediary compounds has the potential to bind with each of the $N_\text{m}$ heterodimer input species. Given mass action kinetics, the formation of the heterodimer-intermediary complexes is driven by their respective concentrations: 
		\begin{align}
			&K^{\text{exp}}_{ij} = \frac{D_{ij}}{\sum_{k=1}^{N_\text{n}} D_{ik}}\\
			&C_j = \sum_{i=1}^{N_\text{m}} K^{\text{exp}}_{ij}X_i. 
		\end{align}
		Where the concentration of the $j$-th intermediary species that binds with the $i$-th heterodimer is $D_{ij}$, $K^{\text{exp}}_{ij}$ is the interaction input matrix, and $C_j$ counts the total concentration of the $j$-th intermediary species among all complexes.
		Due to inhibiting and self-activating reactions of the input-intermediary complexes, only a single chemical species concentration $\bar{C}_j$ persists at the steady state. The output chemical species is defined by the intermediary components and is selected by the largest concentration of intermediary species in all complexes:
		\begin{align}
			&\bar{C}_j = 
			\begin{cases}
				1 & \text{if} \;j = \underset{i}{\text{argmax}}\left( C_i \right)   \\
				0 & \text{otherwise}
			\end{cases}. 
		\end{align}
		This experimental WTA chemical reaction network provides a candidate system which can  experimentally validate our findings which show ICL can be attained in WTA chemical reaction networks. 
		
		In our WTA framework, based on Ref.~\citenum{genot2012computing} and described in Supplementary Materials Section \ref{SIsec:WTAModel}, inputs $\mathbf{z}$ affect reaction rates $k_r$ of a chemical reaction in a manner that selects a final steady state chemical species $\bar{C}_r$ according to Equation \ref{wtaCr}. 
		Although the exact forms of the experimental WTA implementation of Ref.~\citenum{chen2024synthetic} and theoretical WTA network expressions based on Ref.~\citenum{genot2012computing} differ, they nonetheless follow a similar maximum principle which selects a singular chemical species at the steady state. 
		
		Given their mathematical expressions, we suggest the following mapping between theoretical and experimental variables such that our theoretical findings may be experimentally validated. Starting with a trained theoretical WTA network, we may impose values of the trained variables $\mathbf{K}^{\text{eq}}, \ \mathbf{K}$, and $\boldsymbol{\beta}$ onto their experimental counterpart $\mathbf{K}^{\text{exp}}$
		\begin{equation}
			K^{\text{exp}}_{ij} = \frac{K_{ij}}{\beta_j K^{\text{eq}}_j}
		\end{equation}
		where we swapped the index $r$ for $j$.  
		Specifically, we suggest employing values of $K_{ij}^{\text{exp}}$ obtained from combined theoretical trained variables. This is possible given target $K_{ij}^{\text{exp}}$ values can be obtained by tuning the $D_{ij}$ concentrations. Furthermore, the previous mapping permits experimental and theoretical inputs to be identical $\mathbf{z} = \mathbf{X}$, meaning the context vector $\mathbf{z}$ provides now the values of the concentration of the heterodimer inputs $\mathbf{X}$. This is valid only if the lengths of the context and heterodimer input vectors are the same, and $z_i > 0$ possibly through a variable transform. 
		
		If we now substitute the previous mappings into Equation \ref{wtaratekr}  we identify the mapping
		\begin{equation}
			\frac{k_j}{\beta_j} = C_j.
		\end{equation}
		Although the steady state concentrations in the two formulations are not numerically equal, they will have maxima for the same species indices, from which the final logit outputs can be predicted.

		\section{Supplementary results}
		
		\begin{figure*}[ht!]
			\begin{center}
				\includegraphics[width=\textwidth]{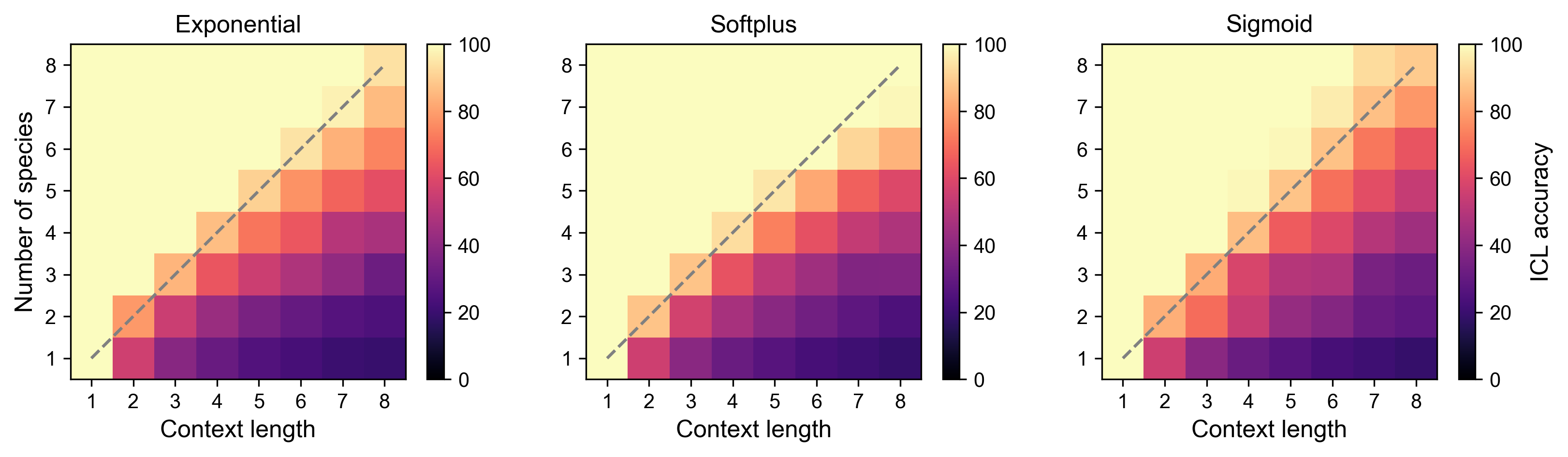}
				\caption{Heat map of the average ICL accuracy over five runs as the context length $N_\text{c}$ and number of species $N_\text{n}$ are varied, using different rate-encoding functions $\sigma$.  Here Softplus refers to $\log\left(1 + \exp(x) \right)$ and Sigmoid refers to $10 / \left(1 + \exp(-x/10)\right)$.
				}
				\label{SI_ActivationFunctions}
			\end{center}
		\end{figure*}
		
		\subsection{ICL is not sensitive to rate-encoding function}\label{SIsec:rateencode}
		
		Throughout this paper we use the rate-encoding function $\sigma = \exp$, although other choices are possible provided that all rates $k_r$ remain positive. Supplementary Figure~\ref{SI_ActivationFunctions} demonstrates that ICL performance is largely insensitive to the specific functional form. Alternative functions, such as Softplus, may offer encoding advantages by introducing a thresholding operation. As our analysis of winner-take-all networks below shows, such thresholding can potentially allow fewer encoding degrees of freedom to achieve high ICL accuracy by assigning default steady-state outcomes when no input projections exceed the threshold. A detailed exploration of how different rate-encoding functions affect encoding and ICL performance is left for future work; we note here only that the main results generalize beyond the use of $\exp$.
		
		\subsection{Analysis of network with two species}\label{SIsec:twospecies}
		To gain intuition for how the pairwise comparison subspaces $\mathcal{M}_i$ can be classified by a chemical reaction network, here we analytically study encoding by small networks.  We first consider linear reaction networks with $N_\text{n} = 2$, $N_\text{c} = 2$, and $D=1$ using the rate-encoding function $\sigma = \text{exp}$.  As discussed in the main text, this network is unable to solve the ICL task, which we explain below.  
		
		The input space is $\mathcal{M} = \mathbb{R}^3$, whose elements' coordinates we denote by $\mathbf{z} = (z_1, z_2, z_q)$ where $q$ stands for ``query.''    We are interested in the two subspaces $\mathcal{M}_1 = \left\{\mathbf{z} \,\big|\, z_1 = z_q\right\}$ and $\mathcal{M}_2 = \left\{\mathbf{z} \,\big|\, z_2 = z_q\right\}$.  These are two planes, with $\mathcal{M}_1$ spanned by the vectors $\mathbf{g}_1 = (1, 0, 1)$ and $\mathbf{e}_2 = (0, 1, 0)$ and $\mathcal{M}_2$ spanned by $\mathbf{e}_1 = (1, 0, 0)$ and $\mathbf{g}_2 = (0, 1, 1)$.
		
		There are two rates in the network, which we denote $k_1$ and $k_2$, defined as learned mappings of the input vector $\mathbf{z}$:
		\begin{equation}
			k_1 = \text{exp}(\mathbf{K}_1^\intercal \mathbf{z}), \quad k_2 = \text{exp}(\mathbf{K}_2^\intercal \mathbf{z})
		\end{equation}
		where we have set $b_r = 0$ for each reaction.  
		The matrix-tree theorem \cite{schnakenberg1976network} for the steady-state vector gives
		\begin{equation}
			\bar{C}_A = \frac{\exp\left(\mathbf{K}_1^\intercal \mathbf{z}\right)}{\exp\left(\mathbf{K}_1^\intercal \mathbf{z}\right) + \exp\left(\mathbf{K}_2^\intercal \mathbf{z}\right)},
		\end{equation}
		with an analogous expression for $\bar{C}_B$, where we have set $C_\text{tot} = 1$.  We see that these functions implement a softmax-like comparison between the projections $\mathbf{K}_1^\intercal \mathbf{z}$ and $\mathbf{K}_2^\intercal \mathbf{z}$.  A sufficient strategy to solve ICL is therefore to learn $\mathbf{K}_1$ and $\mathbf{K}_2$ so that their projections are larger for vectors $\mathbf{z} \in \mathcal{M}_1$ and $\mathbf{z} \in \mathcal{M}_2$ respectively (though this ordering is arbitrary and can be reversed through the decoding matrix $\mathbf{B}$).
		
		To position the vectors $\mathbf{K}_1$ and $\mathbf{K}_2$ so that they best dominate the respective projections with $\mathbf{z} \in \mathcal{M}_1$ and $\mathbf{z} \in \mathcal{M}_2$, they should lie within their corresponding subspaces while being as far as possible from the opposite subspace.  Specifically, we formulate the optimization task as (for the encoding choice just described) $\mathbf{K}_1 \in \mathcal{M}_1$ and $\mathbf{K}_1 \perp \mathcal{M}_{1,2}$, where $\mathcal{M}_{1,2}  \equiv \mathcal{M}_1 \cap \mathcal{M}_2 = \left\{\mathbf{z} \,\big|\, z_1 = z_2 = z_q\right\}$ is the subspace shared by $\mathcal{M}_1$ and $\mathcal{M}_2$.  
		
		To derive the constraints from these two conditions, we first define $\mathbf{K}_1 = (K_{1,1}, K_{1,2}, K_{1,q})$.  To lie within $\mathcal{M}_1$ we require $K_{1,1} = K_{1,q}$.  To be orthogonal to $\mathcal{M}_{1,2}$, which is spanned by $(1,1,1)$, we require $K_{1,2} = -2 K_{1,q}$, so that $\mathbf{K}_1 = K_{1,q}(1,-2,1)$.  We similarly have $\mathbf{K}_2 = K_{2,q}(-2,1,1)$.  The scalars $K_{1,q}$ and $K_{2,q}$ are unconstrained but should be chosen to be large enough in magnitude to saturate their corresponding probability responses.  
		
		For a context vector $\mathbf{z}\in\mathcal{M}_1$, we then have the projections $\mathbf{K}_1^\intercal \mathbf{z} = 2K_{1,q}(z_q - z_2)$ and $\mathbf{K}_2^\intercal \mathbf{z} = -K_{2,q}(z_q - z_2)$.  This reveals a problem: the sign of the projections depends on the relative magnitudes of $z_2$ and $z_q$, yet we only have one unconstrained scalar $K_{1,q}$ to cover both cases.  If we choose $K_{1,q} > 0$ so that $\mathbf{K}_1^\intercal \mathbf{z} > 0$ when $z_q > z_2$, then when $z_q<z_2$ we will have $\mathbf{K}_1^\intercal\mathbf{z} < 0$.  We would require two vectors for each subspace to cover the two sign conditions, amounting to 12 needed degrees of freedom.  With the six available degrees of freedom we are thus unable to span all relevant regions of $\mathcal{M}_1$ and $\mathcal{M}_2$ unless we restrict to input vectors in which, for example, $z_q$ always matches the larger of the two context items.  In the main text we show that this extra restriction on the input data allows the ICL task to be solved.
		
		For the case of arbitrary $D>1$, the same constraints apply as above leading to $\mathbf{K}_1 = \mathbf{K}_{1,q} \odot (1,-2,1)$ (interpreted as an element-wise multiplication) and $\mathbf{K}_2 = \mathbf{K}_{2,q} \odot (-2,1,1)$ where $\mathbf{K}_{1,q}$ and $\mathbf{K}_{2,q}$ play the role of the unconstrained scalars in the case $D=1$.  The dot product with an element $\mathbf{z}\in \mathcal{M}_1$ then reads $\mathbf{K}_1^\intercal \mathbf{z} = 2\mathbf{K}_{1,q}(\mathbf{z}_q - \mathbf{z}_2)$ and $\mathbf{K}_2^\intercal \mathbf{z} = -\mathbf{K}_{2,q}(\mathbf{z}_q - \mathbf{z}_2)$.  As before, the sign of these dot products will depend on the relative directions of $\mathbf{z}_q - \mathbf{z}_2$, $\mathbf{K}_{1,q}$, and $\mathbf{K}_{2,q}$, and no one fixed choice for the unconstrained vectors $\mathbf{K}_{1,q}$ and $\mathbf{K}_{1,q}$ will cover all cases.  
		
		\begin{figure*}[ht!]
			\begin{center}
				\includegraphics[width=1\textwidth]{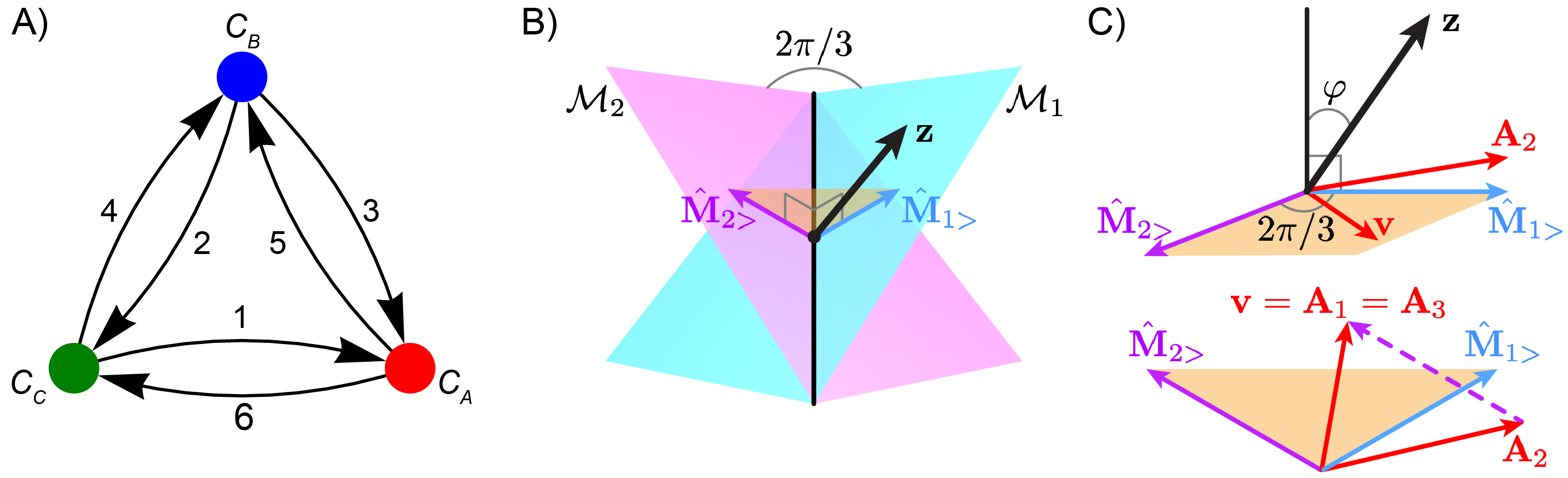}
				\caption{\textbf{Geometry of context space in the three-species network.}
					(A) Diagram of the three-species network with labeled edge rates.
					(B) Schematic of the pairwise comparison subspaces $\mathcal{M}_1$ and $\mathcal{M}_2$, showing the data vector $\mathbf{z}$ and the directions $\hat{\mathbf{M}}_{1>}$ and $\hat{\mathbf{M}}_{2>}$.
					(C) Additional geometric details, including the vectors $\mathbf{A}_1$, $\mathbf{A}_2$, and $\mathbf{A}_3$ arising from the encoding scheme defined by Eqs.~\ref{eqA1}--\ref{eqA3}. The constraint $\mathbf{A}_2 = \mathbf{v} - \mathbf{M}_{2>}$ is illustrated for a general choice of $\mathbf{v}$; we ultimately set $\mathbf{v} = \mathbf{M}_{1>} + \mathbf{M}_{2>}$.}
				\label{SI_3DNetworkGeometry}
			\end{center}
		\end{figure*}

		\begin{figure*}
			\begin{center}
				\includegraphics[width=\textwidth]{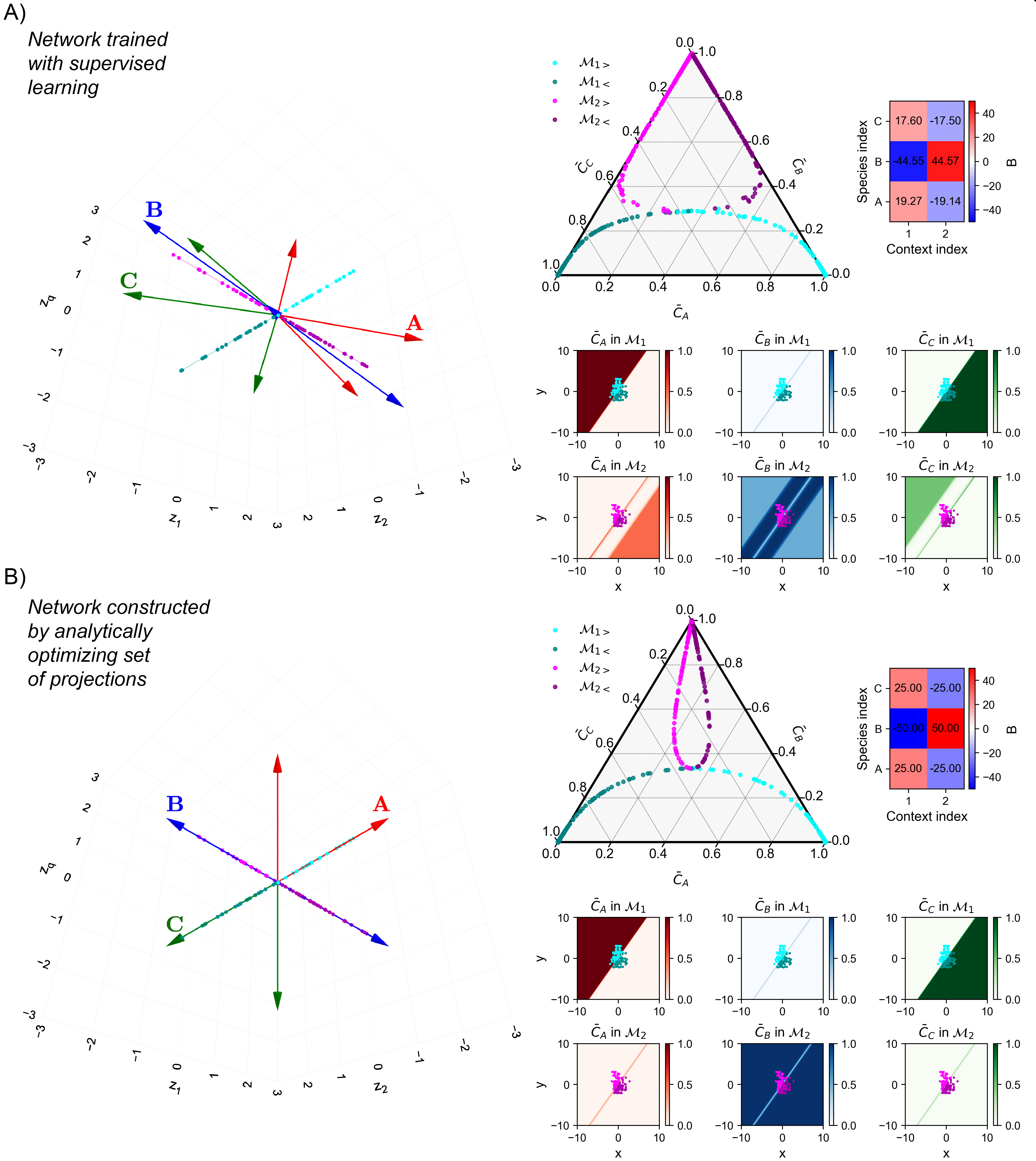}
				\caption{\textbf{Comparison of trained and analytically constructed three-species networks.}
					(A) Parameters and output of a network trained by supervised learning. The 3D plot is rendered as in Fig.~\ref{SmallNetworks} of the main text, with the viewing direction along the shared subspace $\mathcal{M}_1 \cap \mathcal{M}_2$. The ternary plot shows the steady state as in the same figure. The transposed decoding matrix $\mathbf{B}$ is also shown, illustrating how the steady-state vector $\bar{\mathbf{C}}$ is decoded into ICL class predictions $\mathbf{q} = \mathbf{B} \bar{\mathbf{C}}$. In the bottom right, the components of $\bar{\mathbf{C}}$ are plotted over the extended regions of the pairwise comparison planes $\mathcal{M}_1$ and $\mathcal{M}_2$; the diagonal lines reflect intersections with the opposite subspace.
					(B) Same as panel (A), but for the network constructed analytically as described in Supplementary Section~\ref{SIsec:threespecies}.}
				\label{SI_TrainedConstructed}
			\end{center}
		\end{figure*}

		\subsection{Analysis of network with three species}\label{SIsec:threespecies}
		
		We now consider the case with an additional species, i.e., $N_\text{n} = 3$. The input space remains the same, but the network has a larger set of encoding vectors to form projections with the input. There are six rates $k_r$ in the network, which has three spanning trees, with a steady-state expression
		\begin{equation}
			\bar{C}_A = \frac{\sum_{m=1}^{3}\exp\left(\mathbf{A}_m^\intercal\mathbf{z} \right) }{\sum_{m=1}^{3}\exp\left(\mathbf{A}_m^\intercal \mathbf{z} \right) + \sum_{m=1}^{3} \exp\left(\mathbf{B}_m^\intercal\mathbf{z} \right) + \sum_{m=1}^{3} \exp\left(\mathbf{C}_m^\intercal \mathbf{z} \right)}
		\end{equation}
		with analogous expressions for $\bar{C}_B$ and $\bar{C}_C$ using the $\mathbf{B}_m$ and $\mathbf{C}_m$ factors. Using the labeling of species and edge rates in Supplementary Figure~\ref{SI_3DNetworkGeometry}, the spanning-tree vectors are functions of the $\mathbf{K}_r$:
		\begin{eqnarray}
			\mathbf{A}_1 &=& \mathbf{K}_1 + \mathbf{K}_2 \nonumber \\
			\mathbf{A}_2 &=& \mathbf{K}_3 + \mathbf{K}_4 \nonumber \\
			\mathbf{A}_3 &=& \mathbf{K}_1 + \mathbf{K}_3 \nonumber \\
			\mathbf{B}_1 &=& \mathbf{K}_5 + \mathbf{K}_1 \nonumber \\
			\mathbf{B}_2 &=& \mathbf{K}_4 + \mathbf{K}_6 \nonumber \\
			\mathbf{B}_3 &=& \mathbf{K}_4 + \mathbf{K}_5 \nonumber \\
			\mathbf{C}_1 &=& \mathbf{K}_6 + \mathbf{K}_3 \nonumber \\
			\mathbf{C}_2 &=& \mathbf{K}_5 + \mathbf{K}_2 \nonumber \\
			\mathbf{C}_3 &=& \mathbf{K}_6 + \mathbf{K}_2. \nonumber
		\end{eqnarray}
		In total there are 18 degrees of freedom in the $\mathbf{K}_r$ vectors, spread across these 27 functions; thus, not all elements are independent. However, there are sufficient degrees of freedom to allocate two vectors for the subspace $\mathcal{M}_1$ (one for each sign condition of $z_2$ vs $z_q$) and two for $\mathcal{M}_2$.  
		
		We now show how the $\mathbf{K}_r$ vectors can be set to solve the ICL task, and demonstrate that a supervised-learning-trained network approximates this analytical solution. The goal is to place four of the spanning-tree vectors along the branch directions $\hat{\mathbf{M}}_{1>}$, $\hat{\mathbf{M}}_{1<} = - \hat{\mathbf{M}}_{1>}$, $\hat{\mathbf{M}}_{2>}$, and $\hat{\mathbf{M}}_{2<} = - \hat{\mathbf{M}}_{2>}$. These vectors lie within their respective pairwise comparison subspaces $\mathcal{M}_1$ and $\mathcal{M}_2$ and are orthogonal to the intersection $\mathcal{M}_1 \cap \mathcal{M}_2$. They are coplanar and separated by an angle $2\pi/3$ (Supplementary Figure~\ref{SI_3DNetworkGeometry}B). Oppositely signed pairs cover both possible signs of the input's projection.
		
		To simplify the calculation, we choose a specific encoding out of the degenerate set of solutions. Only four vectors are needed to cover the four branch directions, allowing us to zero two of the $\mathbf{K}_r$ vectors. We set $\mathbf{K}_2 = \mathbf{K}_3 = \mathbf{0}$. We assign $\mathbf{B}_1$ and $\mathbf{B}_2$ to the directions $\hat{\mathbf{M}}_{2>}$ and $\hat{\mathbf{M}}_{2<}$ and set $\mathbf{B}_3 = \mathbf{0}$. We fix $\mathbf{B}_1 = \hat{\mathbf{M}}_{2>}$ and $\mathbf{B}_2 = \hat{\mathbf{M}}_{1>}$, normalizing their lengths, though all vectors can be isotropically scaled.
		
		With these choices, five of the $\mathbf{K}_r$ vectors are determined, leaving $\mathbf{K}_1$ as the remaining independent vector, relabeled as $\mathbf{v}$. The relations are then
		\begin{eqnarray}
			\mathbf{K}_1 &=& \mathbf{v} \nonumber \\
			\mathbf{K}_2 &=& \mathbf{0} \nonumber \\
			\mathbf{K}_3 &=& \mathbf{0} \nonumber \\
			\mathbf{K}_4 &=&  -\hat{\mathbf{M}}_{2>} + \mathbf{v} \nonumber \\
			\mathbf{K}_5 &=& \hat{\mathbf{M}}_{2>} - \mathbf{v} \nonumber \\
			\mathbf{K}_6 &=& -\mathbf{v}. \nonumber 
		\end{eqnarray}
		
		Having already covered the directions $\hat{\mathbf{M}}_{2>}$ and $\hat{\mathbf{M}}_{2<}$, the next step is to choose $\mathbf{v}$ so that the directions $\hat{\mathbf{M}}_{1>}$ and $\hat{\mathbf{M}}_{1<}$ are also properly represented, while ensuring that the remaining vectors do not interfere with $\mathbf{B}_1$ and $\mathbf{B}_2$.  The remaining spanning-tree vectors are
		\begin{eqnarray}
			\mathbf{A}_1 &=& \mathbf{v} \label{eqA1} \\
			\mathbf{A}_2 &=& -\hat{\mathbf{M}}_{2>} + \mathbf{v}  \label{eqA2} \\
			\mathbf{A}_3 &=& \mathbf{v}  \label{eqA3} \\
			\mathbf{C}_1 &=& -\mathbf{v} \nonumber \\
			\mathbf{C}_2 &=& \hat{\mathbf{M}}_{2>} - \mathbf{v} \nonumber \\
			\mathbf{C}_3 &=& -\mathbf{v} \nonumber.
		\end{eqnarray}
		
		As illustrated in Supplementary Figure~\ref{SI_3DNetworkGeometry}C, by choosing $\mathbf{A}_1 = \mathbf{A}_3 = \mathbf{v}$ along the direction $\hat{\mathbf{M}}_{1>} + \hat{\mathbf{M}}_{2>}$, the vector $\mathbf{A}_2$ aligns along $\hat{\mathbf{M}}_{2>}$, while $\mathbf{A}_1 = \mathbf{A}_3$ bisects the angle between $\hat{\mathbf{M}}_{1>}$ and $\hat{\mathbf{M}}_{2>}$. This symmetrical choice ensures that $\mathbf{A}_2 = \hat{\mathbf{M}}_{2>}$ and $\mathbf{C}_2 = \hat{\mathbf{M}}_{2<} = -\hat{\mathbf{M}}_{2>}$, while the remaining vectors $\mathbf{A}_1$, $\mathbf{A}_3$, $\mathbf{C}_1$, and $\mathbf{C}_3$ occupy the remaining positions along the hexagonal geometry in the plane spanned by $\hat{\mathbf{M}}_{1>}$ and $\hat{\mathbf{M}}_{2>}$ (see Supplementary Figure~\ref{SI_TrainedConstructed}B).
		
		To verify that this arrangement successfully solves the ICL task, we consider the projections of an input vector $\mathbf{z}$ onto the learned vectors. Using the hexagonal geometry shown in Supplementary Figure~\ref{SI_3DNetworkGeometry}C and a spherical coordinate system with azimuthal angle $\theta = 0$ along $\hat{\mathbf{M}}_{1>}$, we note that all non-zero spanning-tree vectors have equal norms and lie in the same plane. Therefore, the differences in projections between $\mathbf{z}$ and these vectors depend only on the relative azimuthal angles. In this configuration, $\mathbf{A}_2$ has the smallest relative azimuthal angle (zero), and thus dominates the projection for any $\mathbf{z} \in \mathcal{M}_1$, except when $\phi \in (\pi, 2\pi)$, in which case $\mathbf{C}_2$, covering $\mathcal{M}_{1<}$, has the dominant projection.
		
		Finally, Supplementary Figure~\ref{SI_TrainedConstructed} shows that a network trained using supervised learning approximately recapitulates this analytically derived encoding scheme. The trained network achieves high ICL accuracy for inputs sampled from the same range as the training data but fails for inputs drawn from a larger range, indicating limited generalization. This limitation arises because the learned vectors deviate slightly from the ideal solution, causing the projections to reorder when the input range is extended. In contrast, the analytically constructed network maintains the correct projection ordering for any numerical range.

		\begin{figure*}[ht!]
			\begin{center}
				\includegraphics[width=\textwidth]{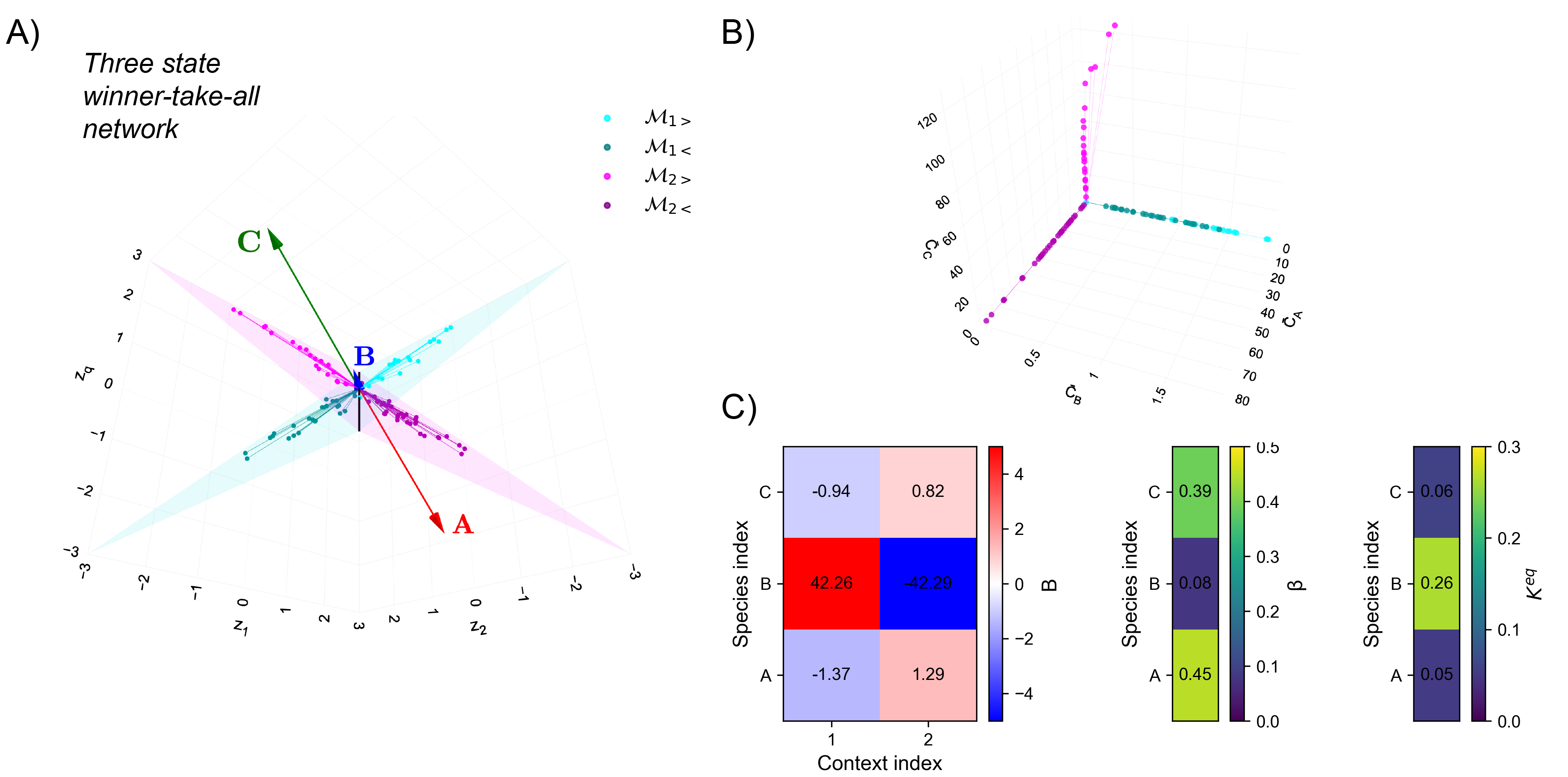}
				\caption{  Illustration of the encoding and decoding strategy in a trained three-species winner-take-all network.  
					A) Learned vectors visualized in the context space.  
					B) Steady-state outputs $\bar{\mathbf{C}}$ for input vectors from different ICL classes.  As there is no conservation among these three species, a three-dimensional representation (rather than a ternary plot) is used.  
					C) Decoding matrix $\mathbf{B}$ along with the learned parameters $\beta$ and $K^\text{eq}$ for each species.
				}
				\label{SI_WTA}
			\end{center}
		\end{figure*}
		
		\begin{figure*}[ht!]
			\begin{center}
				\includegraphics[width=\textwidth]{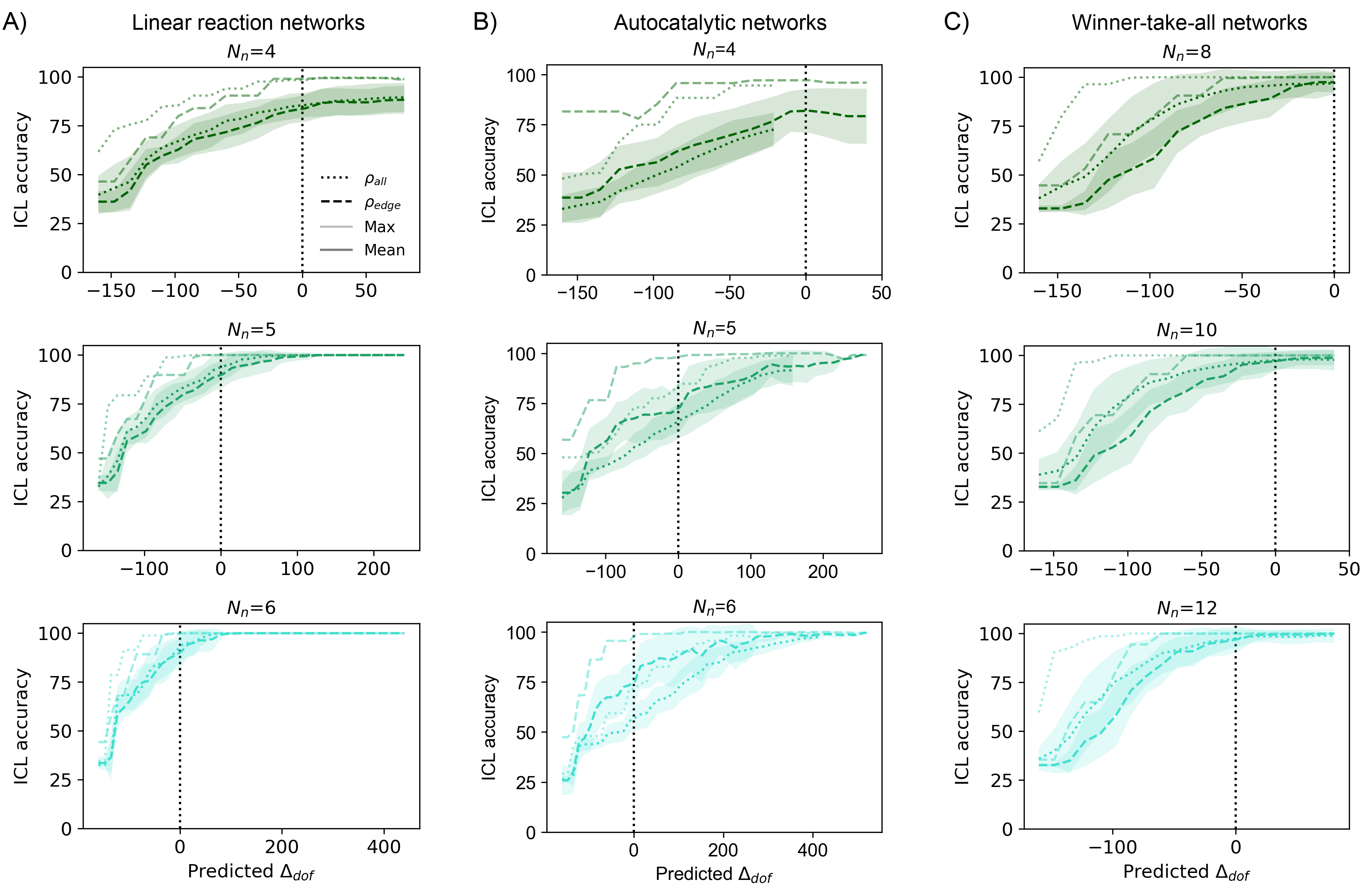}
				\caption{  Additional data corresponding to the experiment shown in Figure~\ref{SparsityResults} of the main text.  
					These plots display the mean, standard deviation, and maximum of ICL accuracy over moving windows of size 50 as a function of the predicted $\Delta_\text{dof}$ (see main text for details).  
					For clarity, data are shown separately for each number of species.  In all plots, dotted lines correspond to sparsity patterns using $\rho_\text{all}$, while dashed lines correspond to sparsity patterns using $\rho_\text{edge}$.  The maximum is indicated by a lighter line, with the mean shown as a solid line.  Panels A-C present data for linear reaction networks, autocatalytic networks, and winner-take-all networks, respectively.
				}
				\label{SI_EdgeAll}
			\end{center}
		\end{figure*}
		
		\subsection{ICL accuracy as a function of encoding sparsity}\label{SIsec:sparsity}
		In Supplementary Figure \ref{SI_EdgeAll} we show additional data from the experiment in Figure \ref{SparsityResults} of the main text.  
		
		\subsection{Analysis of winner-take-all network}\label{SIsec:wta}
		
		Here we analyze the encoding strategy of a winner-take-all (WTA) network with three species to illustrate how its thresholding operation enables the use of fewer degrees of freedom than required in a linear reaction network. We use $N_\text{c} = 2$ and $D = 1$ to allow visualization of the context space in three dimensions.  
		
		Following the setup in Ref.~\citenum{genot2012computing}, we use Equations~\ref{wtaratekr}-\ref{wtaCr} to compute the steady state as a function of the input vector $\mathbf{z}$. With three species, we have three vectors $\mathbf{K}_r$ labeled $\mathbf{A}$, $\mathbf{B}$, and $\mathbf{C}$, corresponding to the species whose production they modulate. These vectors should be positioned in the context space to produce decodable outputs when input vectors lie in different pairwise comparison subspaces $\mathcal{M}_1$ and $\mathcal{M}_2$.  
		
		In Supplementary Figure~\ref{SI_WTA}A, we show how a trained network has positioned the vectors $\mathbf{A}$, $\mathbf{B}$, and $\mathbf{C}$. The vectors $\mathbf{A}$ and $\mathbf{C}$ cover the ICL classes $\mathcal{M}_{2<}$ and $\mathcal{M}_{2>}$; when an input is sampled from these classes, the rates $k_A$ or $k_C$ are large, and the corresponding species $A$ or $C$ dominate the winner-take-all competition via the Softmin operation in Equation~\ref{wtaCr}.  
		
		Interestingly, the learned vector $\mathbf{B}$ is approximately zero and cannot directly cover the remaining ICL classes $\mathcal{M}_{1>}$ and $\mathcal{M}_{1<}$. However, species $B$ still dominates the winner-take-all competition when data is sampled from these classes due to the trained parameters $\beta_B$ and $K^\text{eq}_B$, which are respectively much smaller and larger than for species $A$ and $C$. This gives a default advantage to species $B$, which is only overcome when the rates $k_A$ or $k_C$ are sufficiently large (i.e., when input data is sampled from $\mathcal{M}_{2<}$ or $\mathcal{M}_{2>}$). The $\beta$ and $K^\text{eq}$ parameters thus allow the network to set species $B$ as the default winner through a threshold, and only when data is drawn from $\mathcal{M}_2$ is the threshold exceeded. In this way, the network effectively uses only three learnable vectors to cover the context space.
		
	\end{onecolumngrid}
	
\end{document}